\documentclass[useAMS,usenatbib,a4paper]{mn2e}
\usepackage{amssymb,amsmath}
\usepackage{graphicx}
\usepackage{natbib}
\usepackage{journal_shortcuts}
\newcommand{\sech}{\mathrm{sech}}
\newcommand{\Sun}{_{\sun}}
\newcommand{\Rot}{_{\mathrm{rot}}}
\newcommand{\Therm}{_{\mathrm{therm}}}
\newcommand{\Max}{_{\mathrm{max}}}
\newcommand{\Min}{_{\mathrm{min}}}
\newcommand{\Mean}{_{\mathrm{mean}}}
\newcommand{\Pot}{_{\mathrm{pot}}}
\newcommand{\Kin}{_{\mathrm{kin}}}
\newcommand{\Gal}{_{\mathrm{gal}}}
\newcommand{\ICM}{_{\mathrm{ICM}}}
\newcommand{\DM}{_{\mathrm{DM}}}
\newcommand{\Gas}{_{\mathrm{gas}}}
\newcommand{\Stars}{_*}
\newcommand{\Bulge}{_{\mathrm{bulge}}}
\newcommand{\Ram}{_{\mathrm{ram}}}
\newcommand{\Strip}{_{\mathrm{strip}}}
\newcommand{\Bound}{_{\mathrm{bnd}}}
\newcommand{\Grav}{_{\mathrm{grav}}}
\newcommand{\Disc}{_{\mathrm{disc}}}
\newcommand{\degree}{^o}
\newcommand{\K}{\,\textrm{K}}
\newcommand{\Kpc}{\,\textrm{kpc}}
\newcommand{\PC}{\,\textrm{pc}}
\newcommand{\Yr}{\,\textrm{yr}}
\newcommand{\Myr}{\,\textrm{Myr}}
\newcommand{\Gyr}{\,\textrm{Gyr}}
\newcommand{\Kms}{\,\textrm{km}\,\textrm{s}^{-1}}
\newcommand{\Cm}{\,\textrm{cm}}
\newcommand{\Erg}{\,\textrm{erg}}
\newcommand{\ccm}{\,\textrm{cm}^{-3}}
\newcommand{\gccm}{\,\textrm{g}\,\textrm{cm}^{-3}}

\title[Ram pressure stripping of inclined disc galaxies]%
{Ram pressure stripping of disc galaxies:\newline The role of the inclination angle}

\author[E. Roediger and M. Br\"uggen]%
{Elke Roediger%
\thanks{E-mail:
e.roediger@iu-bremen.de (ER); m.brueggen@iu-bremen.de (MB)}
and  
Marcus Br\"uggen\footnotemark[1]%
\\
International University Bremen, P.O. Box 750\,561, 28725 Bremen,
Germany}

\begin{document}

\date{Accepted. Received; in original form }

\pagerange{\pageref{firstpage}--\pageref{lastpage}} \pubyear{2005}

\maketitle

\label{firstpage}

\begin{abstract}
We present 3D hydrodynamical simulations of ram pressure stripping of massive
disc galaxies in clusters. Studies of galaxies that move face-on have
predicted that in such a geometry the galaxy can lose a substantial amount of
its interstellar medium. But only a small fraction of galaxies is moving
face-on. Therefore, in this work we focus on a systematic study of the effect
of the inclination angle between the direction of motion and the galaxy's
rotation axis.

In agreement with some previous works, we find that the inclination angle does
not play a major role for the mass loss as long as the galaxy is not moving
close to edge-on (inclination angle $\lesssim 60\degree$). We can predict this
behaviour by extending Gunn~\& Gott's estimate of the stripping radius, which
is valid for face-on geometries, to moderate inclinations.

The inclination plays a role as long as the ram pressure is comparable to
pressures in the galactic plane, which can span two orders of magnitude. For
very strong ram pressures, the disc will be stripped completely, and for very
weak ram pressures, mass loss is negligible independent of inclination. We
show that in non-edge-on geometries the stripping proceeds remarkably
similar. A major difference between different inclinations is the degree of
asymmetry introduced in the remaining gas disc.

We demonstrate that the tail of gas stripped from the galaxy does not
necessarily point in a direction opposite to the galaxy's direction of
motion. Therefore, the observation of a galaxy's gas tail may be misleading
about the galaxy's direction of motion.
\end{abstract}


%
%
%
%
%
\section{Introduction}
\label{sec:intro}
Galaxies populate different environments, ranging from isolated field galaxies
to dense galaxy clusters. Comparative observations between field and cluster
galaxies reveal that, in addition to internal processes, the environment plays
an important role in galaxy evolution. Disc galaxies are especially
affected. Many well-known differences between cluster and field disc galaxy
populations are attributed to a smaller gas content of cluster members. Such
differences are the higher fraction of red galaxies and early-type galaxies in
clusters (e.g.~\citealt{goto03b,pimbblet03}) as well as the HI-deficiency of
cluster disc galaxies (\citealt{cayatte90,cayatte94,solanes01}) and the reduced
star formation rates
(\citealt{koopmann98,koopmann04b,koopmann04a,koopmann05}).

Several processes have been proposed to explain these features. One
idea are tidal interactions between cluster members. Although the interaction
timescales in clusters are rather short due to the high velocity dispersion of
cluster galaxies, Moore et al. (1996,1998,1999)
\nocite{moore96,moore98,moore99} have shown that repeated short interactions
(harassment) can have a substantial influence on cluster galaxies (see also
review by \citealt{mihos04} and references therein). Another process is ram
pressure stripping, which is the interaction between the galactic gas disc and
the intracluster medium (ICM). As galaxies move through the ICM, the ram
pressure is expected to push out (parts of) their interstellar medium (ISM),
provided the ram pressure is strong enough. This process does not affect the
dynamics of the stellar and dark matter components of the galaxy. If ram
pressure stripping plays a role, one expects to find galaxies with undisturbed
stellar discs and distorted and truncated gaseous discs. Such examples have
been observed e.g.~NGC 4522
(\citealt{kenney99,kenney01,kenney04},\citealt{vollmer04a}), NGC 4548
(\citealt{vollmer99}) and NGC 4848 (\citealt{vollmer01}).

The ICM-ISM interaction is a complex process influenced by many
parameters. Different aspects have been studied by several
groups. \citet{schulz01}, \citet{marcolini03} and \citet{roediger05} showed
that the ICM-ISM interaction is a multi-stage process. The most important
phases are the instantaneous stripping, on a timescale of a few $10\Myr$, an
intermediate phase, on a timescale of up to a few $100\Myr$, and a continuous
stripping phase that, in principle, could continue until all gas is lost from
the galaxy. Instantaneous stripping is the immediate consequence of the ram
pressure dislocating the gas disc. However, the gas does not become unbound
from the galactic potential immediately. It takes the intermediate phase until
all displaced gas is truly lost. \citet{schulz01} described this as a
``hang-up'' phase where gas lingers behind the disc. The continuous stripping
is caused by the Kelvin-Helmholtz (KH) instability induced by the ICM wind
flowing over the surface of the remaining gas disc, leading to a slow but
continuous gas loss. This process is also termed turbulent/viscous stripping
(see e.g.~\citealt{nulsen82,quilis00}).

A list of parameters that influence ram pressure stripping includes the ICM
wind density, $\rho\ICM$, wind velocity, $v\ICM$, the inclination angle, $i$,
between the disc's rotation axis and the wind direction, the overall galaxy
mass (its potential depth) and its gas mass, the variability of the ICM wind,
the structure of the ISM gas disc. \citet{gunn72} proposed that for galaxies
moving face-on through the ICM the success or failure of ram pressure
stripping can be predicted by comparing the ram pressure with the galactic
gravitational restoring force per unit area. If this comparison is done for
each galactic radius, a stripping radius can be estimated. In general,
numerical simulations agree with this analytical estimate. For inclined
geometries, no clear picture has yet emerged. In SPH simulations,
\citet{abadi99} found a good agreement with the analytical estimate for
face-on cases. For the few inclined cases they studied, their model galaxies
lost significantly less gas. However, this result may be biased by their short
runtime of $\sim 100\Myr$. Using an SPH code, \citet{schulz01} studied the
phases of the ICM-ISM interaction in detail. They focused on face-on
geometries, but also performed a few simulations with inclined galaxies. They
found that inclined galaxies are stripped on longer timescales. They also
studied the loss of angular momentum and found that inclined galaxies are
``annealed'' by a stronger loss of angular momentum. Hence, the galaxies
contract and thus become more resistant against stripping. \citet{quilis00}
used a hydrodynamical grid code to study, both, the influence of the
inclination angle and an inhomogeneous gas disc structure. In their work, they
concentrated on rather high ram pressures, for which they found the
inclination angle to play a minor role as long as the galaxy is not moving
strictly edge-on. So far the work of \citet{vollmer01a} is the only study of
the effect of a time-dependent ram pressure, as it would be the case for
galaxies orbiting in clusters. Using a sticky particle code, they modelled the
ram pressure as an additional variable force on the gas disc particles. They
observed a significant backfall of material that had been stripped during the
time of maximum ram pressure but was not unbound. This is in agreement with
the result of \citet{schulz01} and \citet{roediger05} that even for a constant
ICM wind the unbinding of stripped material takes a few $100\Myr$. The sticky
particle code of \citet{vollmer01a} cannot model hydrodynamical processes such
as the continuous stripping found with hydro-codes because it cannot simulate
hydrodynamical instabilities. Nonetheless, this model has been applied
successfully to individual galaxies
(e.g.~\citealt{vollmer00,vollmer01,vollmer99,vollmer03}). \citet{marcolini03}
studied ram pressure stripping of discy dwarf galaxies in groups, where the
ram pressure is lower than in clusters. They used a hydrodynamical 3D grid
code and studied inclinations of $0\degree$, $45\degree$ and $90\degree$. This
group found that in dwarf galaxies the inclination angle only plays a
role as long as the central disc pressure is comparable to the ram
pressure. While for smaller ram pressures little gas is lost, for stronger ram
pressures the complete disc is stripped. Even in the range where inclination
plays a role, the $45\degree$ runs were similar to the face-on cases and only
edge-on cases could retain more gas. \citet{roediger05} presented a
comprehensive parameter study of face-on stripping. They varied wind density
and velocity and covered conditions from cluster centres to outskirts and
galaxy groups. The ram pressure, $p\Ram = \rho\ICM v\ICM^2$, depends on, both,
ICM density and velocity, and usually it is assumed the stripping depends on
$p\Ram$, and not on $\rho\ICM$ and $v\ICM$ individually. While this was
confirmed in general by \citet{roediger05}, they found that supersonic runs
tend to suffer slightly less than corresponding subsonic runs. This work used
a 2D hydrodynamical grid code and was thus restricted to the face-on geometry.

Even though ram pressure stripping has been studied by several groups, most
simulations were performed for a face-on geometry, either because the code was
2D or because this geometry is expected to show the strongest
impact. Qualitative results for the influence of the inclination angle have
been found, but they do not agree in all points. So here, we concentrate on
the influence of the inclination angle in the case of massive galaxies and
present 3D hydrodynamical simulations that were performed with the
adaptive-mesh-refinement code FLASH. We study the amount of gas loss and
discuss asymmetrical structures introduced in the remaining gas disc. We also
present projected gas surface density maps which reveal interesting
implications for the interpretation of observational data.

In Sect.~\ref{sec:method}, we briefly introduce the code and describe our
galaxy model and simulation parameters. In Sect.~\ref{sec:analytical}, we
discuss some analytical considerations, and in Sect.~\ref{sec:results} we
present our simulation results. Finally, in Sect.~\ref{sec:discussion} we
compare our results to previous work and discuss their implications.

%

\section{Method}
\label{sec:method}
We study the motion of the galaxy through the cluster in the
rest frame of the galaxy. Therefore, its motion translates into an ICM wind
flowing past the galaxy. 

\subsection{Code} \label{sec:code}
The simulations were performed with the FLASH code (\citealt{fryxell00}), a hydrodynamical
adaptive-mesh-refinement code with a PPM hydro solver. All boundaries but
the inflow boundary are open. The simulations presented here are performed in
3D. We use a simulation box of size of $(x\Min,x\Max)\times
(y\Min,y\Max)\times (z\Min,z\Max)=(-64.8\Kpc,64.8\Kpc)^3$. For the low ram
pressure runs a box size of $(x\Min,x\Max)\times
(y\Min,y\Max)\times (z\Min,z\Max)=(-97.2\Kpc,97.2\Kpc)\times (-64.8\Kpc,129.6\Kpc)\times (-97.2\Kpc,97.2\Kpc)$ was used to prevent
interference with the boundaries. The galactic centre is
located at $(x\Gal,y\Gal,z\Gal)=(0,0,0)$. The ICM wind is flowing along the
$y$-axis into the positive direction. Most simulations were done using 4
refinement levels, resulting in an effective resolution of $500\PC$ and an
effective number of grid cells of $256^3$ ($384^3$ for the low ram pressure runs). For comparison, we also run a few
test simulations with 5 refinement levels (effective grid size of $512^3$,
and effective resolution of $250\PC$) (see Appendix~\ref{sec:resolution}). The
refinement criteria were the standard density and pressure gradient criteria.

In the FLASH code, we used a mass scalar to ``dye'' the galactic gas. Thus, we
can identify which fraction of gas inside a grid cell originated from the
galactic disc (for a more detailed description of such ``dyeing'' techniques
see \citealt{roediger05}).

\subsection{Model galaxy}
\begin{figure}
\centering\resizebox{0.7\hsize}{!}{\includegraphics[angle=0]{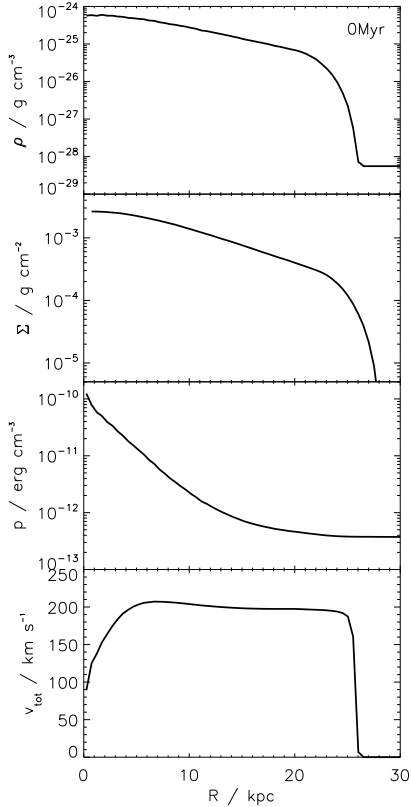}}
\caption{Radial profiles of the density, $\rho$, pressure, $p$,
  and rotation velocity, $v\Rot$, in the galactic plane for the initial
  model. Also the radial profile for the projected ISM surface density,
  $\Sigma$, is shown.}
\label{fig:initial_profiles}
\end{figure}

We model a massive spiral galaxy with a flat rotation curve at $200\Kms$. It
consists of a gas disc, a stellar disc, a stellar bulge and a dark matter (DM)
halo. For the gravitational potential of the stellar disc, bulge and DM halo
we use the following analytical descriptions:
\begin{description}
\item[Stellar disc:] Plummer-Kuzmin disc, see \citet{miyamoto75} or
  \citet{binneytremaine}. Such discs are characterised by their mass,
  $M\Stars$, and radial and vertical scale lengths, $a\Stars$ and $b\Stars$,
  respectively.
\item[Stellar bulge:] Spherical Hernquist bulge (see \citealt{hernquist93}). In
  case of a spherical bulge, the gravitational potential, $\Phi$, depends on
  radius, $r$, as $\Phi(r)= -
  \frac{G\,M\Bulge}{r+r\Bulge}$, where $M\Bulge$ is the mass of the bulge,
  $r\Bulge$ the scale radius and $r$ the spherical radius.
\item[DM halo:] The spherical model of \citet{burkert95}, including the
  self-scaling relations, i.e. the DM halo is characterised by the radial
  scale length $r\DM$ alone. For the equation of the analytical potential see
  also \cite{mori00}.
\end{description}
The self-gravity of the gas is neglected as the gas contributes only a small
fraction to the overall galactic mass. The galaxy model parameters are
summarised in Table~\ref{tab:galaxy_parameters}.
%
\begin{table}
\caption{Galaxy model parameters.}
\label{tab:galaxy_parameters}
\centering\begin{tabular}{lll}
\hline
             &$M\Stars$   & $10^{11}M\Sun$ \\
stellar disc &$a\Stars$   & 4\,kpc                    \\
             &$b\Stars$   & 0.25\,kpc                 \\
\hline
bulge        &$M\Bulge$   & $10^{10}M\Sun$ \\
             &$r\Bulge$   & 0.4\,kpc                  \\
\hline
DM halo      &$r\DM$      & 23\,kpc                   \\
\hline
             &$M\Gas$     & $10^{10}M\Sun$             \\
gas disc     &$a\Gas$     & 7\,kpc                     \\
             &$b\Gas$     & 0.4\,kpc                  \\
             &$v\Rot$     & 200$\Kms$             \\
\hline
\end{tabular}
\end{table}
%
In order to prevent steep density gradients in the galactic plane and in the
galactic centre, the gas disc is described by a softened exponential disc :
\begin{equation}
\rho(R,Z)=\frac{M\Gas}{2\pi a\Gas^2 b\Gas}  \,0.5^2\,\sech\left(\frac{R}{a\Gas}\right)\, \sech\left(\frac{|Z|}{b\Gas}\right) \label{eq:dens_exp_soft}.
\end{equation}
The coordinates $(R,Z)$ are galactic cylindrical coordinates. The radial and
vertical scale lengths are $a\Gas$ and $b\Gas$, respectively. For $R\gtrsim
a\Gas$ and $|Z|\gtrsim b\Gas$, this density distribution converges towards the
usual exponential disc $\rho(R,Z)=\frac{M\Gas}{2\pi a\Gas^2 b\Gas}
\exp(-R/a\Gas)\,\exp(-|Z|/b\Gas)$. For the corresponding exponential disc,
$M\Gas$ is the total gas mass. We chose $M\Gas$ such that in the outer regions
the gas disc converges to an exponential gas disc with 10\% of the stellar
disc mass, i.e.  $M\Gas=0.1 M\Stars$. Due to the reduction of the gas density
in the central part of the softened exponential disc, the integrated gas mass
amounts to $6\cdot 10^{9}M\Sun$. Given the density distribution in the disc,
its pressure and temperature distribution are set such that hydrostatic
equilibrium with the ICM is maintained in the direction perpendicular to the
disc plane. In radial direction, the disc's rotation velocity is set so that
the centrifugal force balances the gravitational force and pressure
gradients. We have cut the gas disc smoothly to a finite radius of $26\Kpc$ to
prevent interaction of stripped gas with the grid boundaries. This smooth
cutting was achieved by multiplying the gas density distribution $\rho(R,Z)$
with $0.5(1+ \cos(\pi(R-20\Kpc)/6\Kpc ) )$ for $20\Kpc < R \le
26\Kpc$. Figure~\ref{fig:initial_profiles} shows radial profiles of density,
surface density, pressure and rotation velocity for the initial model.
%
%

We measure the inclination angle, $i$, of the galaxy as the angle between
its rotation axis and the ICM wind direction. Thus $0\degree$ corresponds to
to a face-on motion of the galaxy, and $90\degree$ to edge-on.

\subsection{ICM conditions}
In order to study the influence of the inclination angle, $i$, isolatedly, we
use a constant ICM wind. The additional effect of a variable ICM wind, as one
would expect for a cluster passage, will be studied in a forthcoming paper.

The strength of the ICM wind is characterised by the ram pressure,
$p\Ram=\rho\ICM v\ICM^2$, which depends on, both, ICM wind density, $\rho\ICM$,
and velocity, $v\ICM$. \citet{roediger05} have shown that the ram pressure is
the relevant parameter and not $\rho\ICM$ or $v\ICM$. Therefore, we use
$v\ICM=800\Kms$ for most simulations and vary the ram pressure by varying
$\rho\ICM$. We chose the ICM temperature such that the sound speed is
$1000\Kms$. Thus, the standard $v\ICM$ corresponds to a Mach number of
0.8. For comparison, a few runs with supersonic $v\ICM=2530\Kms$ have also
been performed.  The parameters for $p\Ram$, $\rho\ICM$, $v\ICM$ and $i$ are
summarised in Table~\ref{tab:wind_parameters}.
%
\begin{table}
\caption{ICM wind parameters. Crosses indicate which ICM wind
  parameter - inclination combinations have been simulated.}
\label{tab:wind_parameters}
\centering\begin{tabular}{lll||lllll}
\hline
&&& \multicolumn{5}{c}{inclination $i/\degree$} \\
$p\Ram/$ & $\rho\ICM/$ & $v\ICM/$ & &&& \\
$(\Erg\ccm)$ & $(\gccm)$ &  $(\Kms)$ & 0 & 30 & 60 & 75 & 90 \\
\hline
$6.4\cdot 10^{-11}$ & $10^{-26}$ & 800 & $\times$ & $\times$ & $\times$ &  & $\times$ \\
$6.4\cdot 10^{-12}$ & $10^{-27}$ & 800 & $\times$ & $\times$ & $\times$ &$\times$ & $\times$ \\
$6.4\cdot 10^{-12}$ & $10^{-28}$ & 2530 &  & $\times$ & & $\times$ & \\
$6.4\cdot 10^{-13}$ & $10^{-28}$ & 800 & $\times$ & $\times$ & $\times$ &  & $\times$ \\
\hline
\end{tabular}
\end{table}
%
We will refer to the three ram pressures used as low, medium and high. If not
stated otherwise, we refer to runs with $v\ICM=800\Kms$.

We start the simulation with the ICM at rest and then increase the inflow
velocity over the first $50\Myr$ from zero to $v\ICM$.
 
In order to include all stages of the ICM-ISM interaction, the overall runtime
of our simulations is $1\Gyr$.

\section{Analytical considerations}
\label{sec:analytical}
%
%
\subsection{Different phases} \label{sec:phases}
As described in previous works, the ICM-ISM interaction proceeds in multiple
phases. The discussion of the phases in Sect.~\ref{sec:intro} made clear that
the overall process is quite complex. Analytical estimates exist mainly for
the instantaneous stripping phase in face-on geometry (see
Sect.~\ref{sec:gunn_gott}). \citet{marcolini03} estimated a stripping radius
for the edge-on geometry by assuming that the instantaneous stripping can be
neglected and the stripping radius will be set by the KH-instability. They
estimated the radius up to which the galaxy's gravity can suppress the
KH-instability, and assumed that gas outside this radius will be
stripped. They found that, both, the estimates for face-on and edge-on cases
are rather similar. However, the timescale for the continuous stripping, which is
the relevant process in the edge-on geometry, operates on a much longer
timescale. \citet{roediger05} estimated that the mass loss rate during the
continuous stripping phase is of the order of $1 M\Sun/\Yr$.

In the following sections, we discuss the estimate for the stripping radius in
the instantaneous stripping phase and try to derive an estimate for inclined
geometries.

\subsection{Face-on geometry} \label{sec:gunn_gott}
The usual way to estimate the amount of gas loss during the instantaneous
stripping phase for galaxies moving face-on through the ICM follows the
suggestion of \citet{gunn72}. Here, one compares the gravitational
restoring force per unit area and the ram pressure for each radius of the
galaxy. At radii where the restoring force is larger, the gas can be retained,
at radii where the ram pressure is larger, the gas will be stripped. The
transition radius is called the stripping radius. In order to calculate the
restoring force per unit area, it is assumed that the gas disc can be
approximated by an infinitely thin disc. Then, the restoring force per unit
area would be
\begin{equation}
f\Grav(R) = \Sigma\Gas(R)\; \frac{\partial\Phi}{\partial Z}(R) \label{eq:rest_force},
\end{equation}
where $\Phi$ is the gravitational potential of the galaxy and $\Sigma\Gas(R)$
the gas surface density. As pointed out by e.g.~\citet{roediger05}, the
gradient of $\partial\Phi/\partial Z$ exactly in the disc plane is
zero, so if $\partial\Phi/\partial Z$ exactly in the galactic plane is
used, this estimate would not yield any restoring force. In order to obtain a
reasonable estimate, one can use the maximum gradient
$\frac{\partial\Phi}{\partial z}(R,Z)$ behind the disc plane at a given $R$. In
general, the resulting stripping radii and retained gas masses from numerical
simulations agree well with the predictions from this simple estimate.

\subsection{Inclined cases} \label{sec:analytic_inclination}
Now we wish to move on from face-on geometries to inclined cases. As in
face-on geometries, we assume that the gas disc can be described by an
infinitely thin disc. For a given surface element d$A$ at radius $R$ in the
galactic plane, the ram pressure force is $\rho\ICM v\ICM^2 \cos i\,$d$A$,
because we have to use the projected cross-section of this surface
element. This force is axisymmetrical with respect to the galactic rotation
axis for all $i$. We can also interpret this force in terms of an {\em
effective} inclination-dependent ram pressure $\cos i\,\rho\ICM v\ICM^2$.
Clearly, the ram pressure force decreases with inclination. In addition to the
reduced ram pressure force, also the correct gravitational restoring force
needs to be calculated. For a surface element d$A$ in the galactic plane at
galactic radius $R$ and azimuthal angle $\phi$, the restoring force is the
projection of the sum of local gravitational force and centrifugal force onto
wind direction. Since the centrifugal force and the radial gravitational force
in a disc galaxy balance, we can drop both terms from our analysis. Only the
projection of the gravitational force perpendicular to the disc plane onto
wind direction remains. This is $\cos\,i\cdot f\Grav \,$d$A$ with $f\Grav$
calculated as in Eqn.~\ref{eq:rest_force}. Again, this force is axisymmetrical
with respect to the galactic rotation axis. If now the ram pressure force and
the restoring force are compared for all disc radii, the factor $\cos i$ drops
out completely. Consequently, the stripping radius should be independent of
inclination. One of the assumptions for this estimate is that the gas disc can
be approximated by an infinitely thin disc. Clearly, this assumption is not
valid for highly inclined cases, so the estimate is expected to fail for
near-edge-on cases. However, for moderate inclinations it could give a
useful approximation. For near-edge-on inclinations, we expect that the galaxy
loses less gas.

\section{Simulation results}
\label{sec:results}
The most striking result of our simulations is that with increasing
inclination a stronger and stronger degree of asymmetry is introduced to the
gas disc. This asymmetry makes simple concepts such as a stripping radius
difficult to apply.

\subsection{Snapshots of the evolution}
\begin{figure*}
\includegraphics[trim=0 80 0 0,clip,angle=0,width=0.32\textwidth]{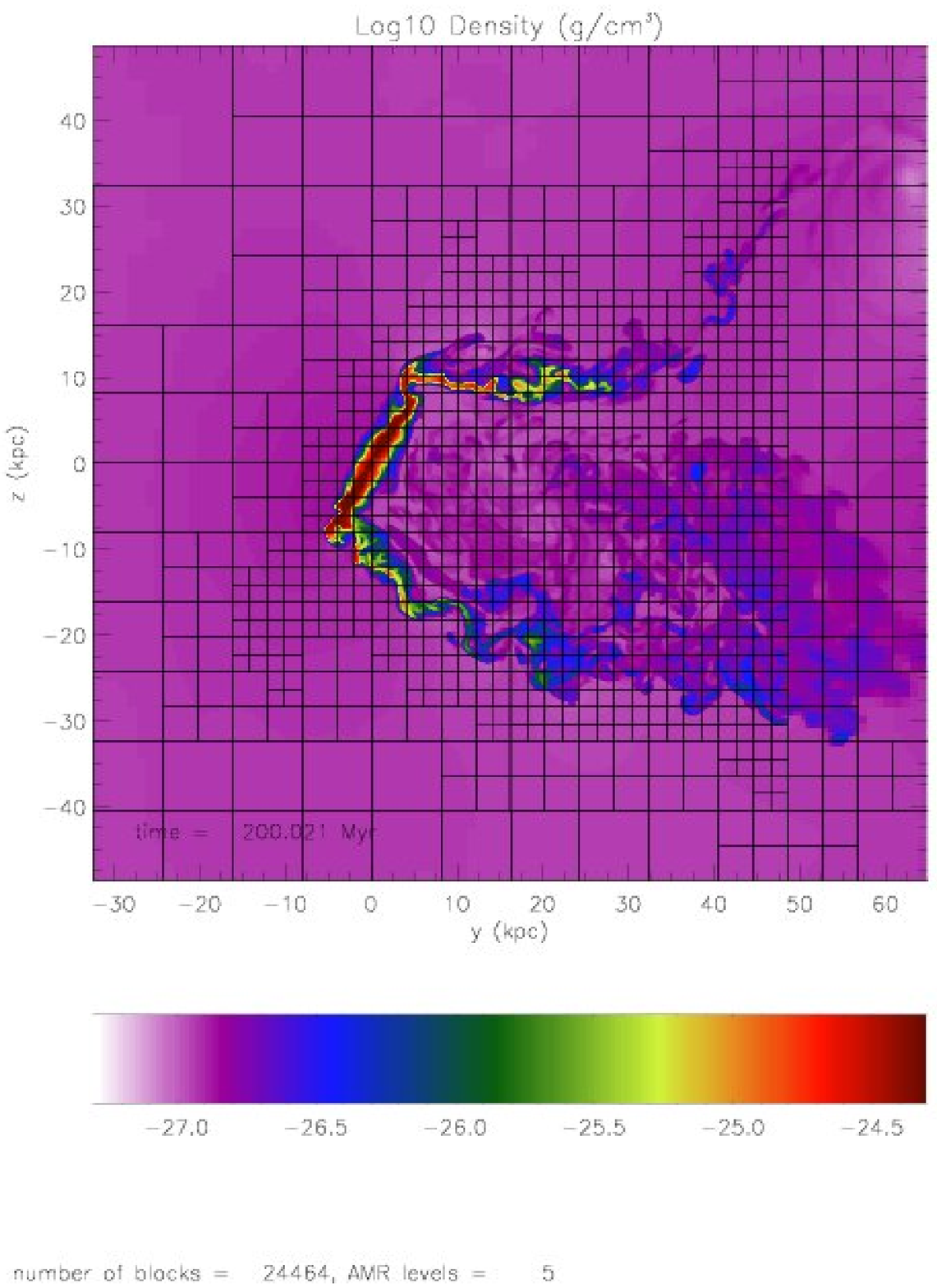}
\includegraphics[trim=0 80 0 0,clip,angle=0,width=0.32\textwidth]{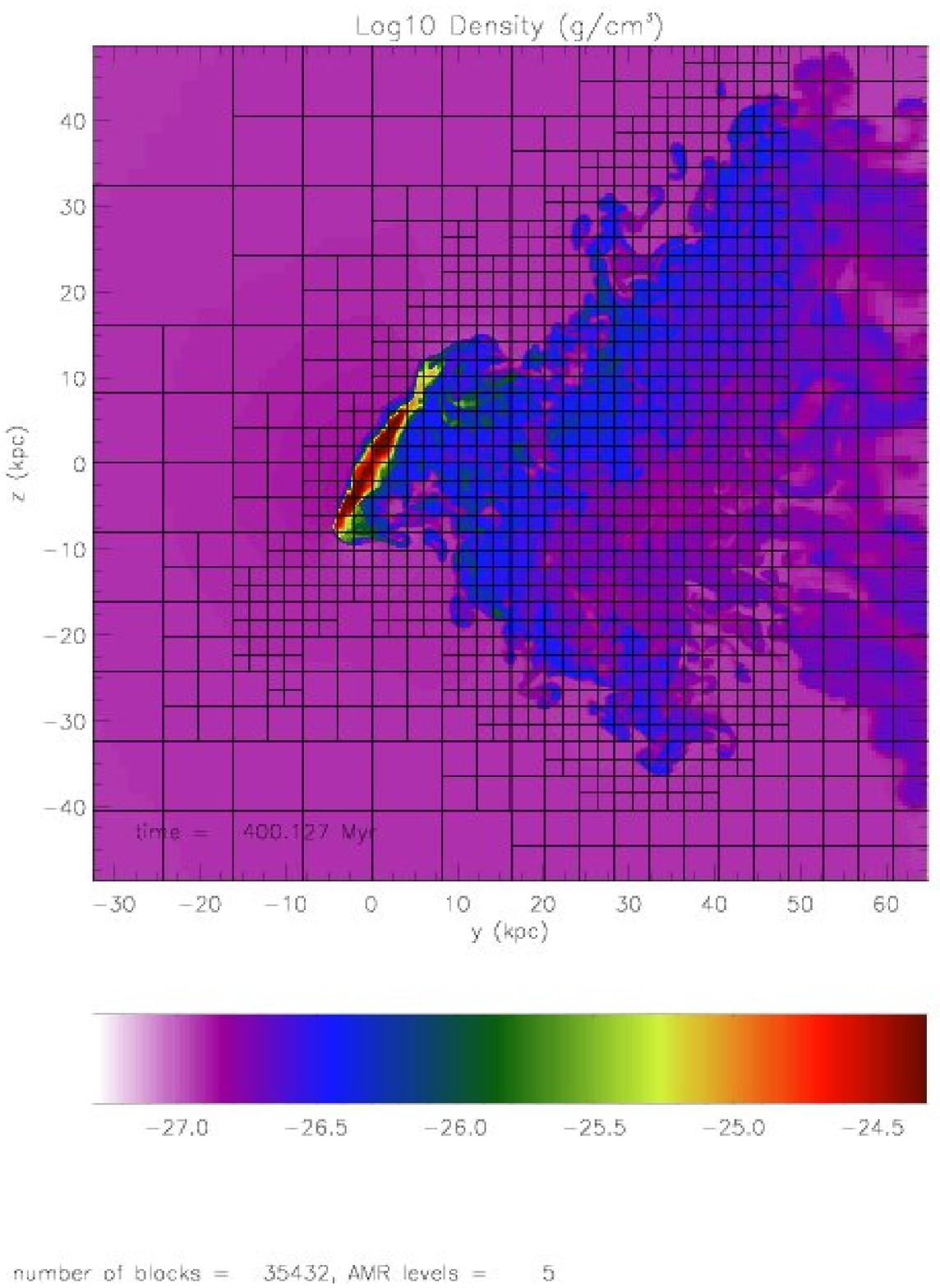}
\includegraphics[trim=0 80 0 0,clip,angle=0,width=0.32\textwidth]{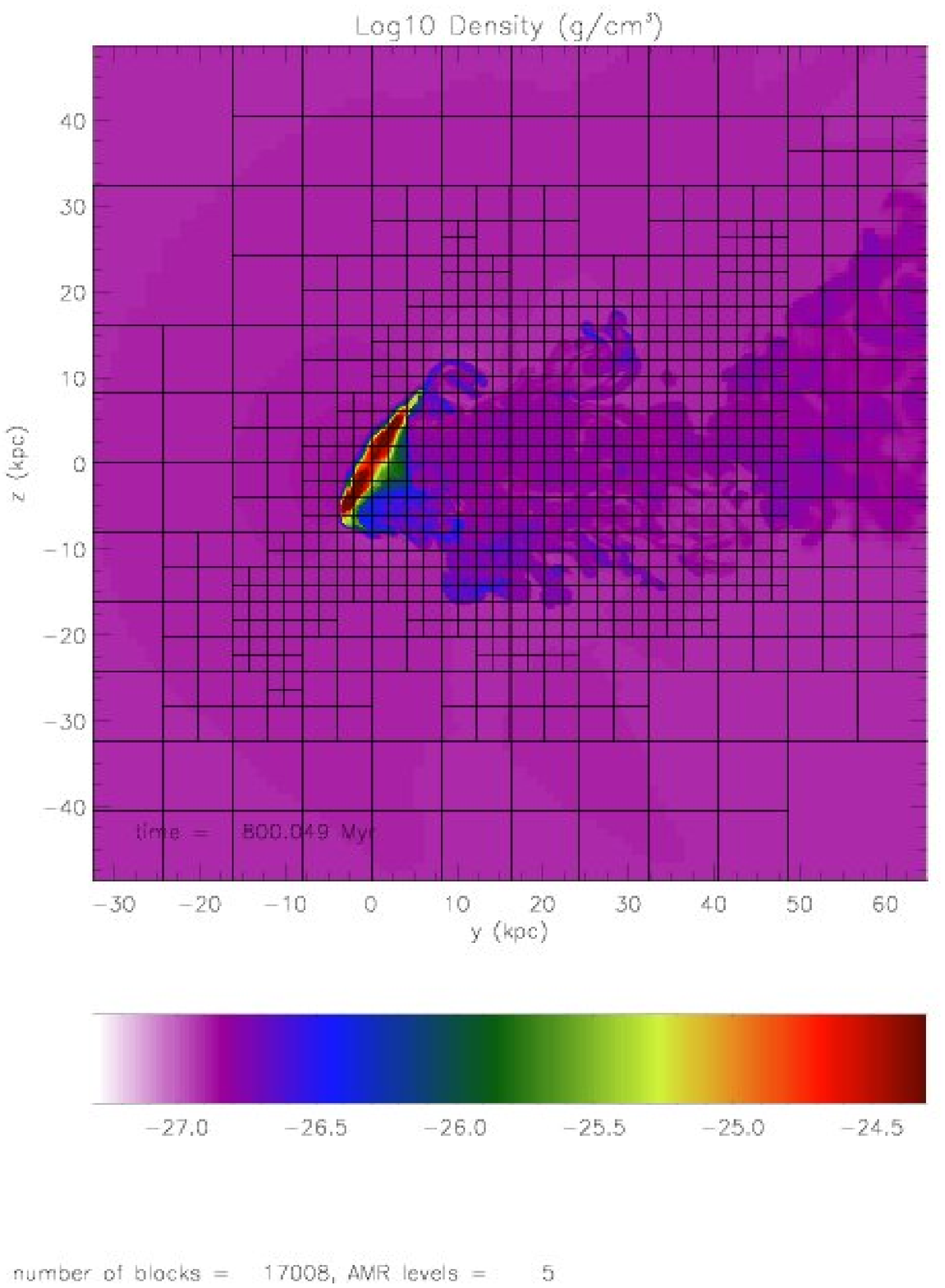}
\caption{Slices in $y$-$z$-plane at $x=0$, colour-coded density for different
  timesteps. For medium ram pressure, inclination
  of $30\degree$. The ICM wind is flowing from left to right. Also the grid
  structure is shown: One square is the crossection of one block$=8^3$ grid
  cells. These snapshots are taken from the run with $250\PC$ resolution.}
\label{fig:slice_i30}
\end{figure*}

\begin{figure*}
\includegraphics[angle=90,trim=50 80 0 0,clip,width=0.32\textwidth]{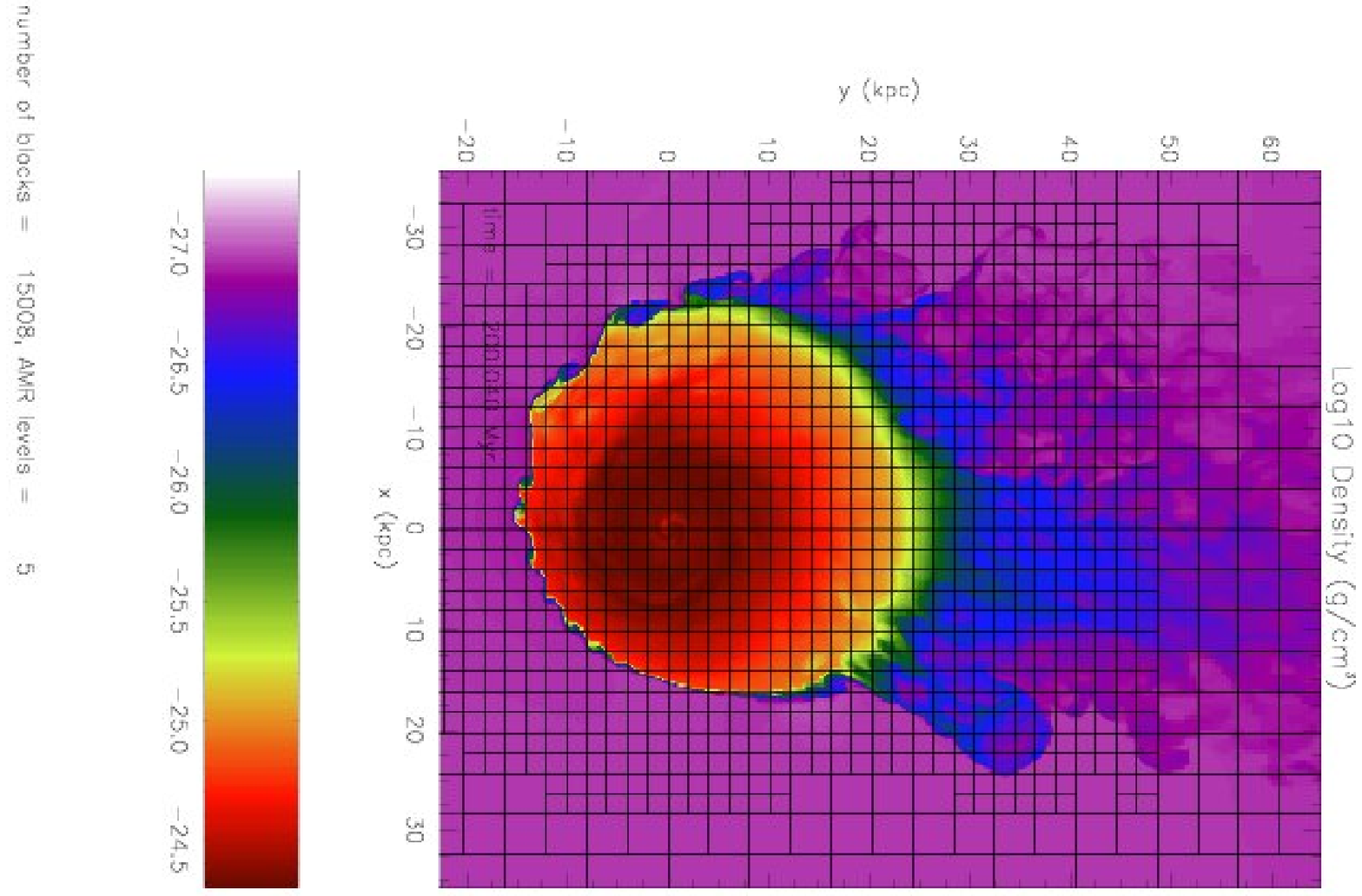}
\includegraphics[angle=90,trim=50 80 0 0,clip,width=0.32\textwidth]{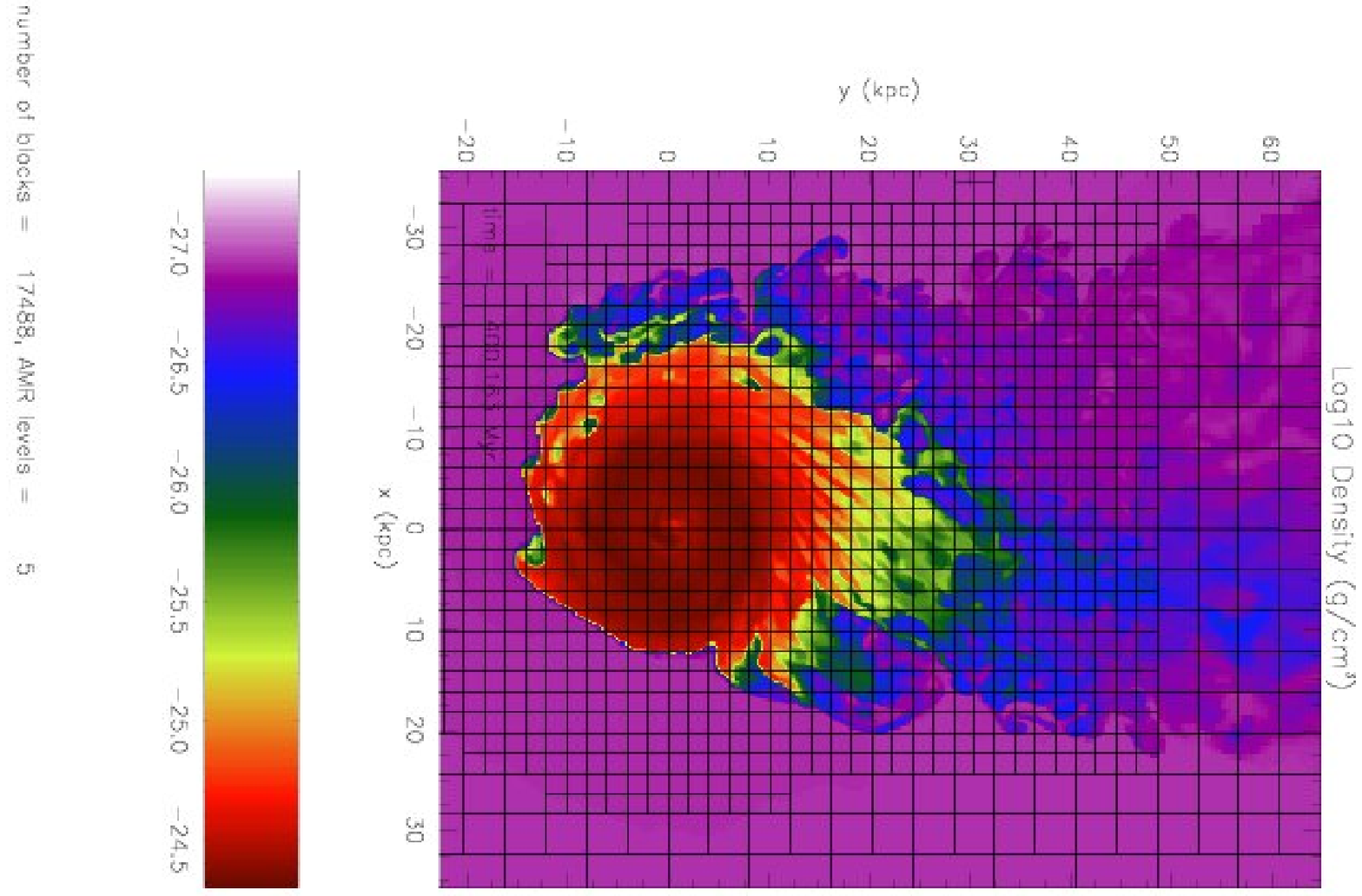}
\includegraphics[angle=90,trim=50 80 0 0,clip,width=0.32\textwidth]{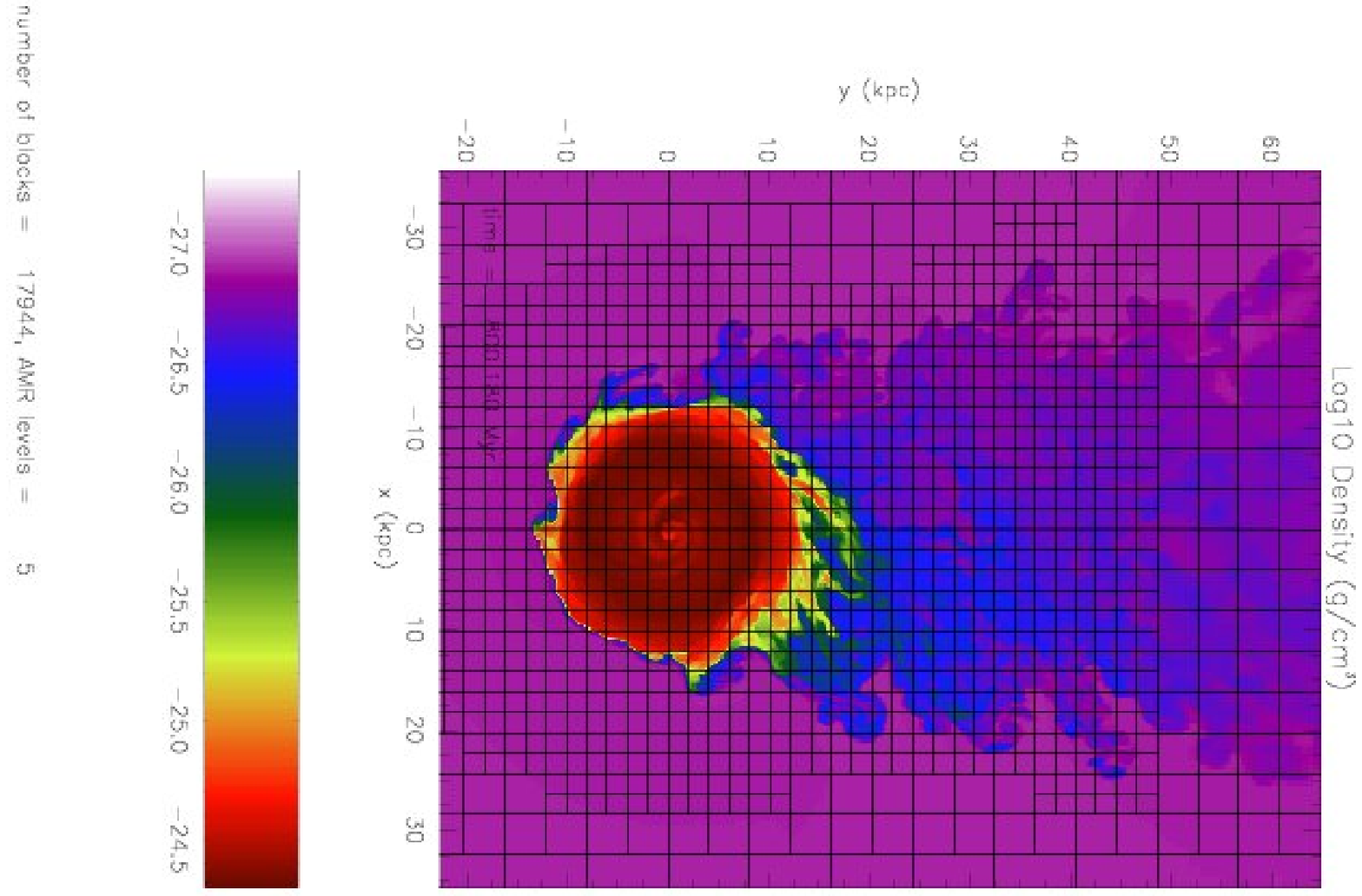}
\caption{Slices in $x$-$y$-plane at $z=0$, colour-coded density for different
  timesteps. For medium ram pressure, inclination
  of $90\degree$. The ICM wind is flowing from bottom to top. For more details
  see Fig.~\ref{fig:slice_i30}.}
\label{fig:slice_i90}
\end{figure*}

%
To visualise the 3D data, we show slices through the simulation box and
projected gas densities. Figures~\ref{fig:slice_i30} and \ref{fig:slice_i90}
display the local gas density in certain planes through the galaxy for an
inclination of $30\degree$ (Fig.~\ref{fig:slice_i30}) and $90\degree$
(Fig.~\ref{fig:slice_i90}). 
%
%
%
Both plots represent runs with medium ram pressure. Snapshots at 3 different
times are shown. For the $30\degree$ case, the disc remains rather symmetrical,
whereas especially early snapshots of the $90\degree$ case display a strong
degree of asymmetry.

\begin{figure*}
\includegraphics[width=0.32\textwidth]{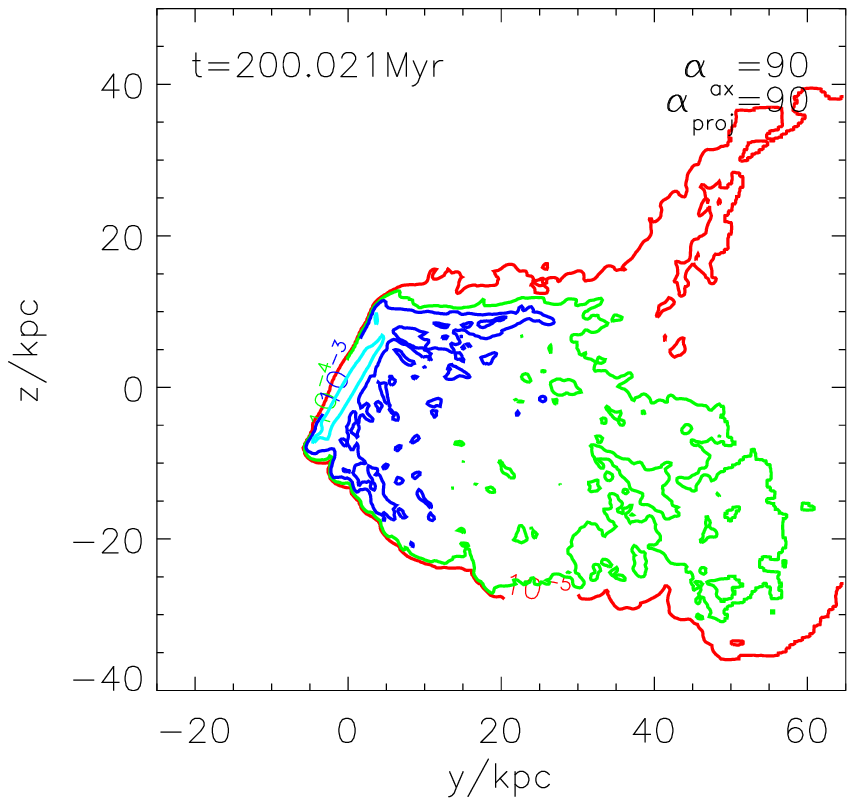}
\includegraphics[width=0.32\textwidth]{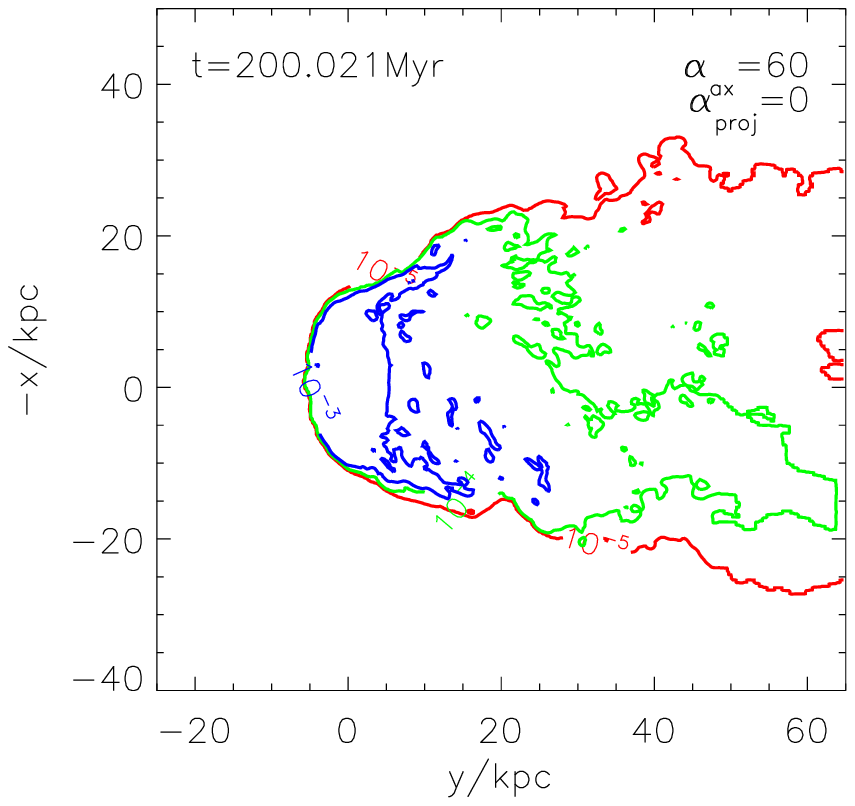}
\includegraphics[width=0.32\textwidth]{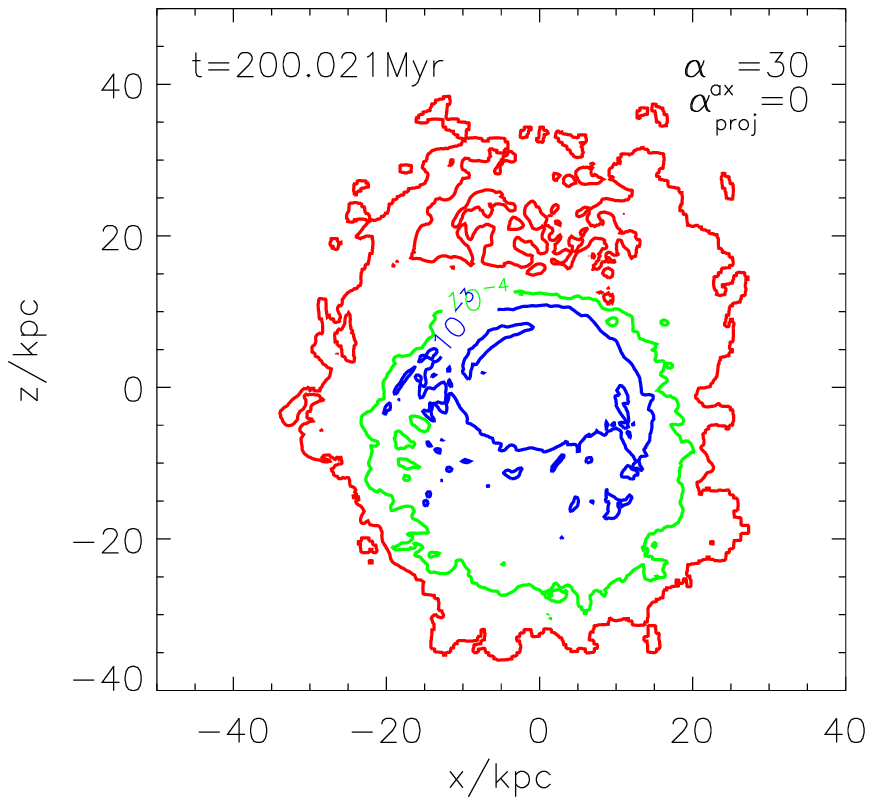}\\
\includegraphics[width=0.32\textwidth]{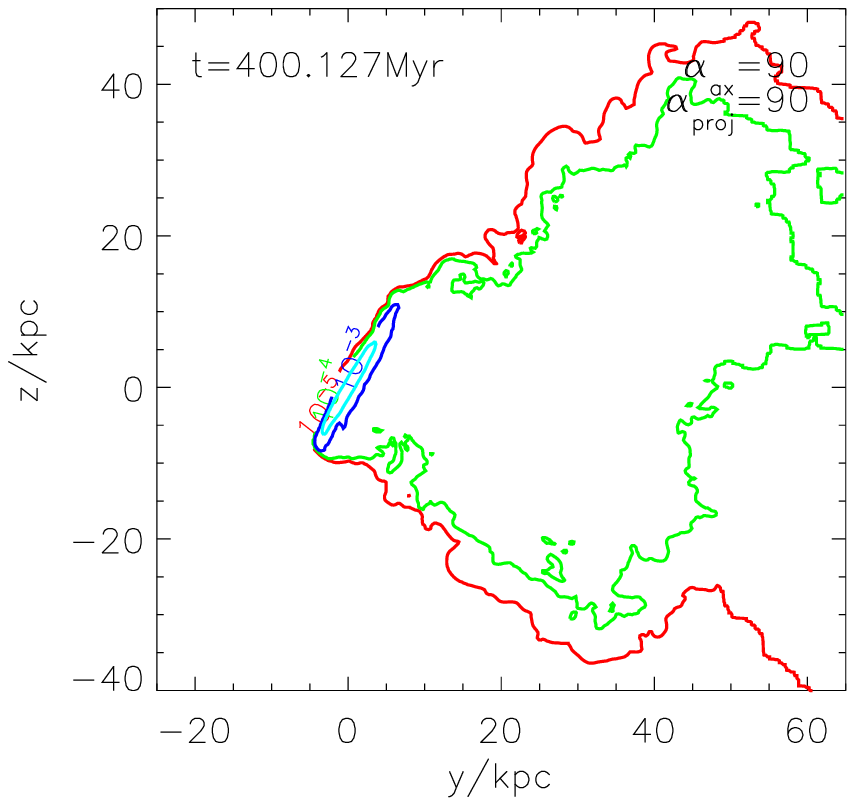}
\includegraphics[width=0.32\textwidth]{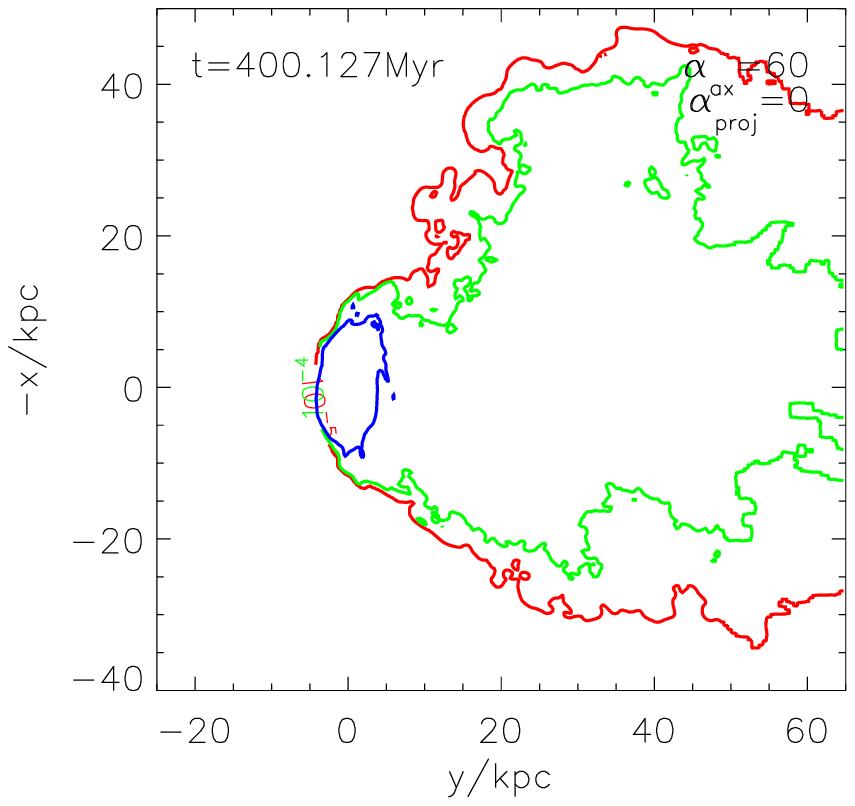}
\includegraphics[width=0.32\textwidth]{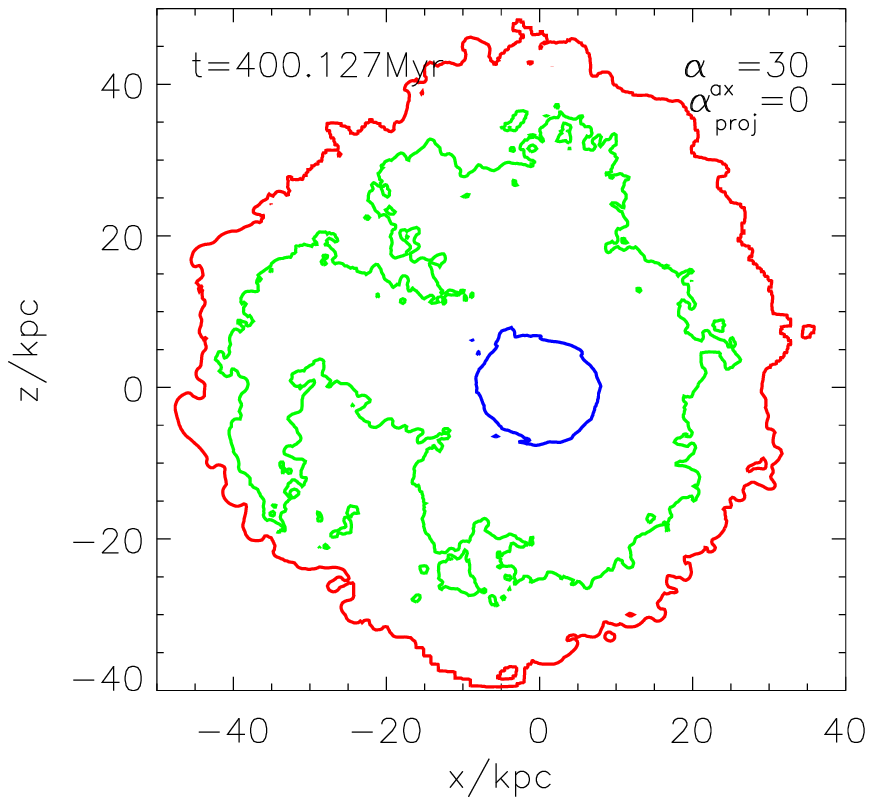}\\
\includegraphics[width=0.32\textwidth]{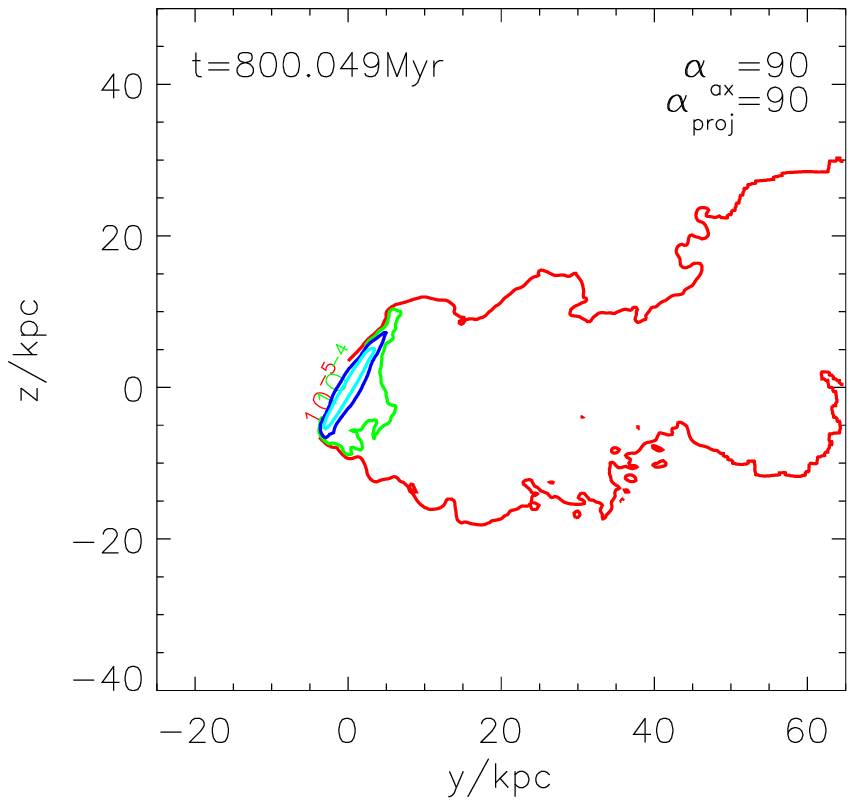}
\includegraphics[width=0.32\textwidth]{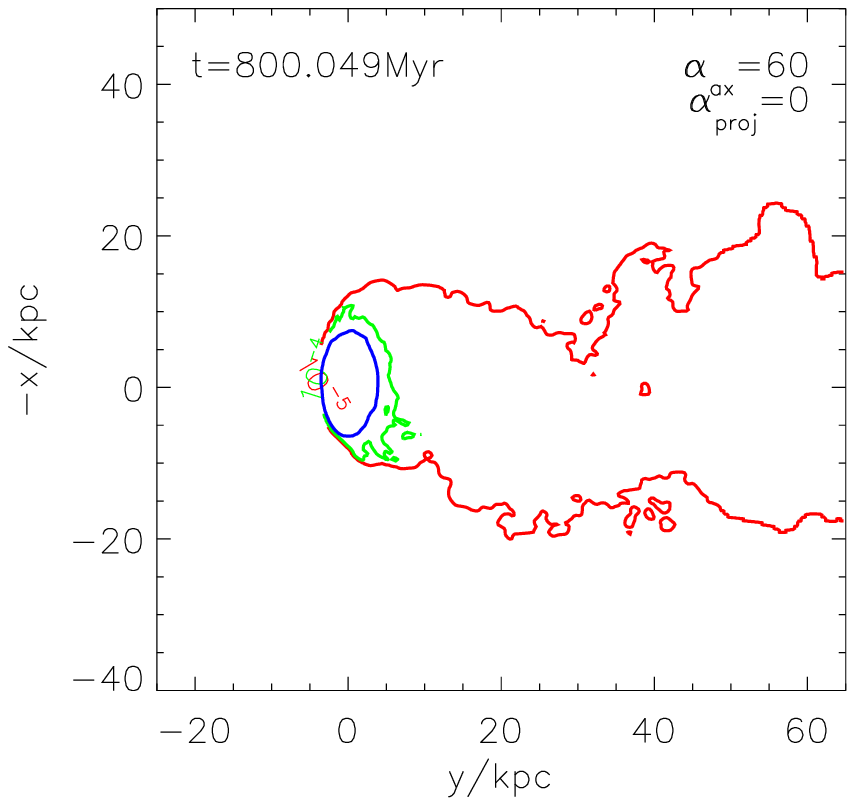}
\includegraphics[width=0.32\textwidth]{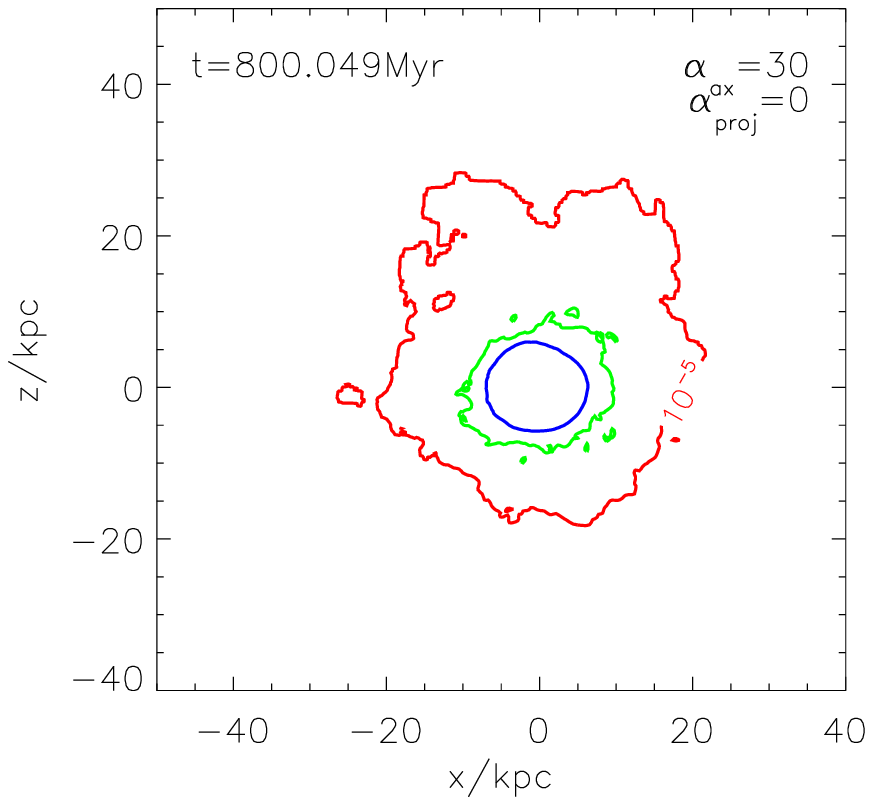}
\caption{Projected gas densities along $x$-, $z$-, and $y$-axis in the left,
  middle and right column, respectively. Each row is for the same timestep
  that is printed in the top right corners of each panel. For medium ram
  pressure and inclination of $30\degree$. Contour levels are $10^{-5,-4,-3,-2}\,\mathrm{g}\,\mathrm{cm}^{-2}$.}
\label{fig:proj_dens_i30}
\end{figure*}

\begin{figure*}
\includegraphics[width=0.32\textwidth]{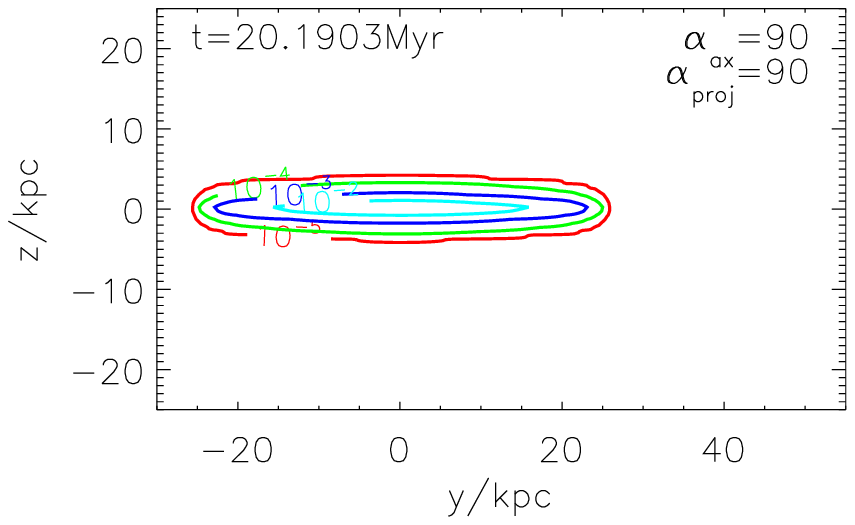}
\includegraphics[width=0.32\textwidth]{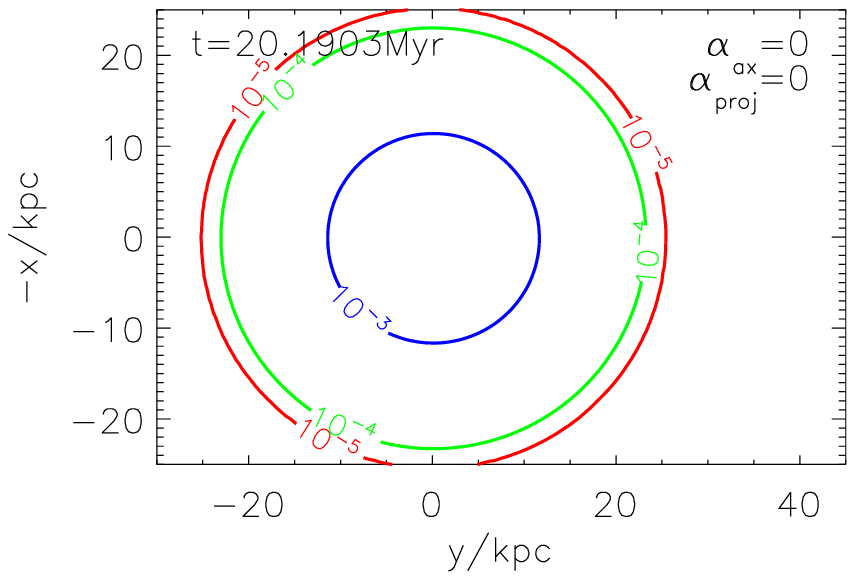}
\includegraphics[width=0.32\textwidth]{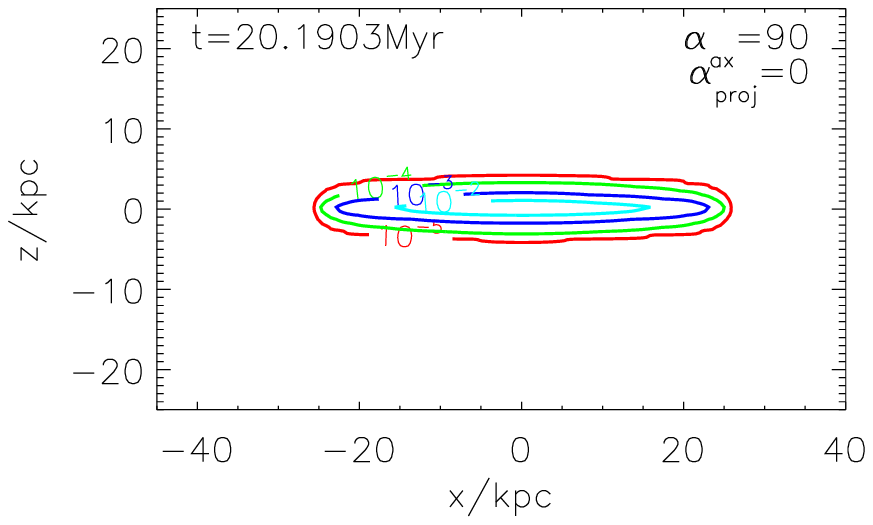}\\
\includegraphics[width=0.32\textwidth]{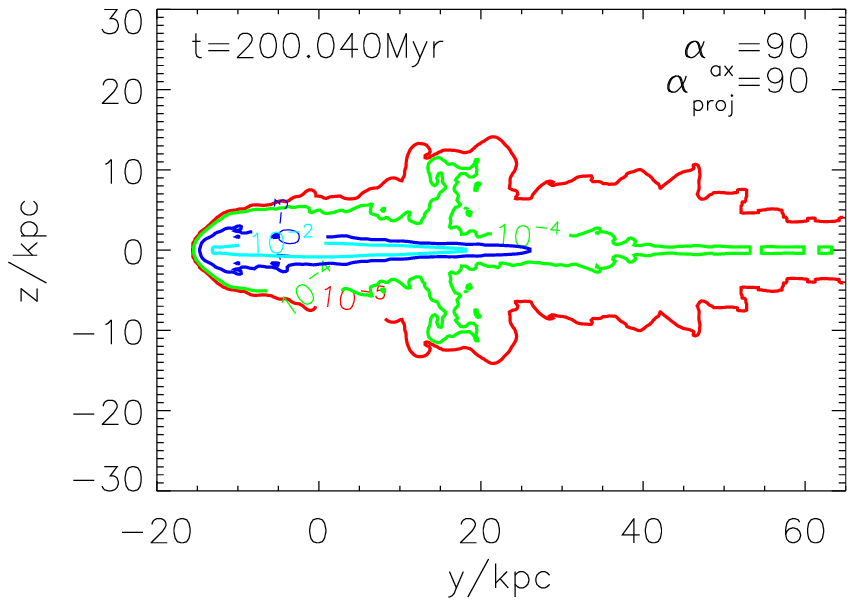}
\includegraphics[width=0.32\textwidth]{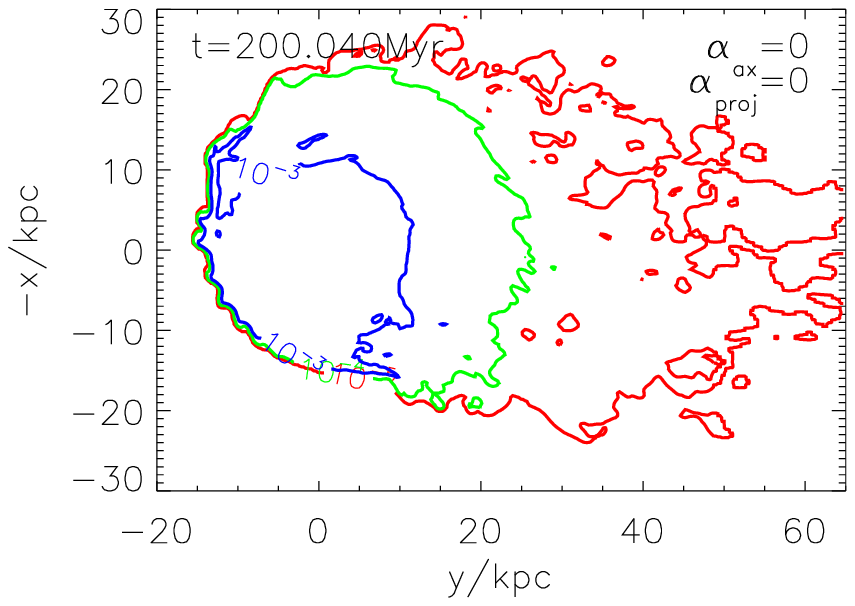}
\includegraphics[width=0.32\textwidth]{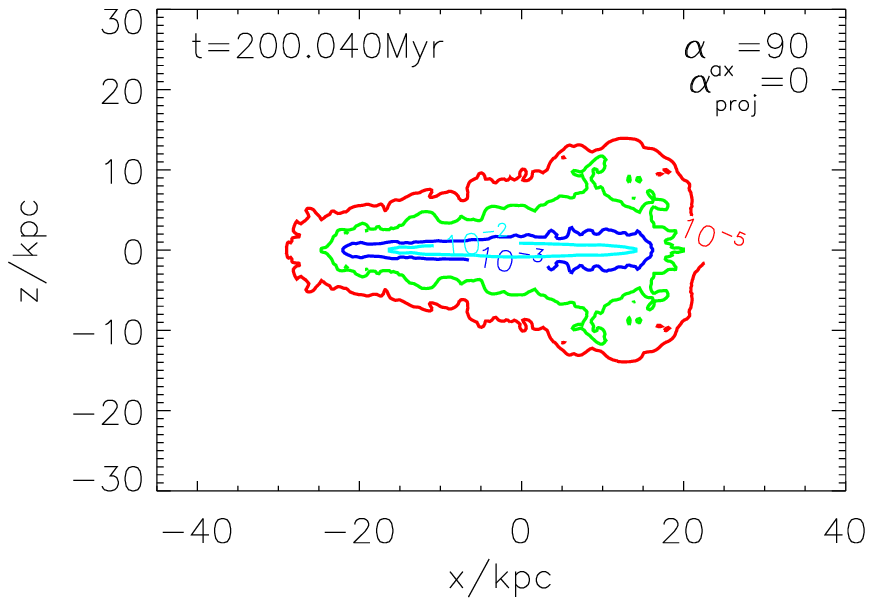}\\
\includegraphics[width=0.32\textwidth]{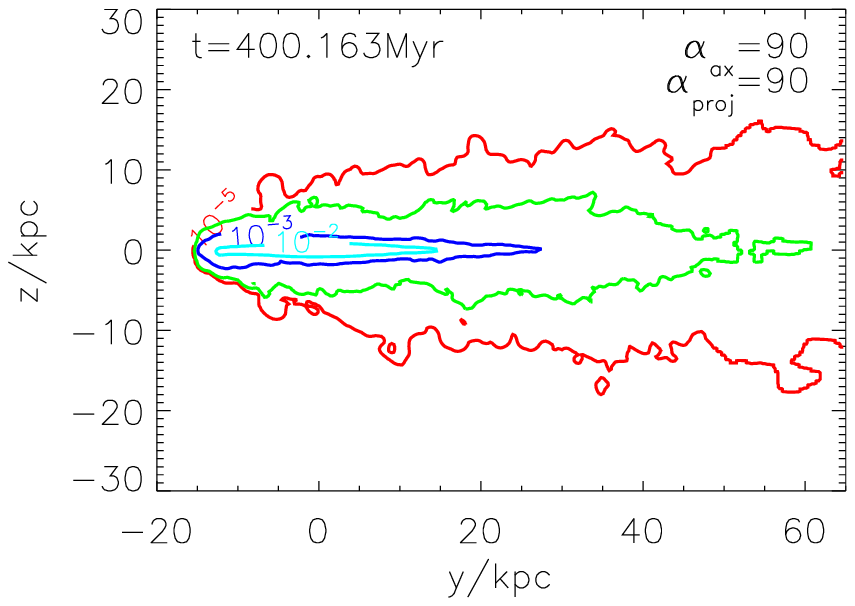}
\includegraphics[width=0.32\textwidth]{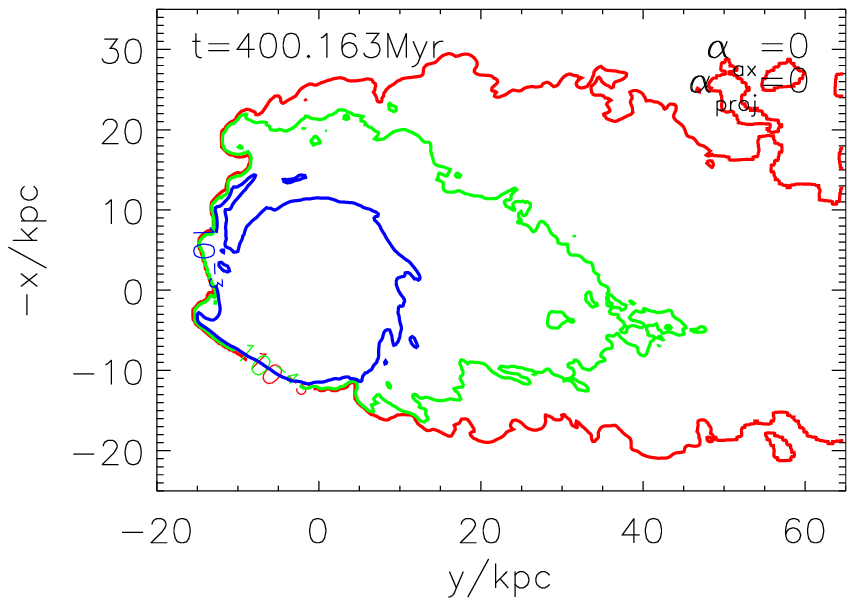}
\includegraphics[width=0.32\textwidth]{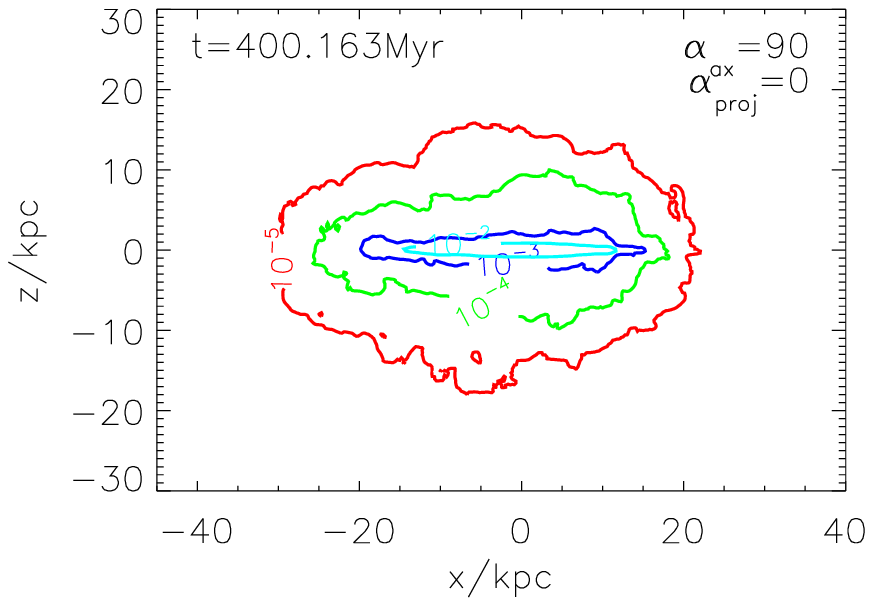}\\
\includegraphics[width=0.32\textwidth]{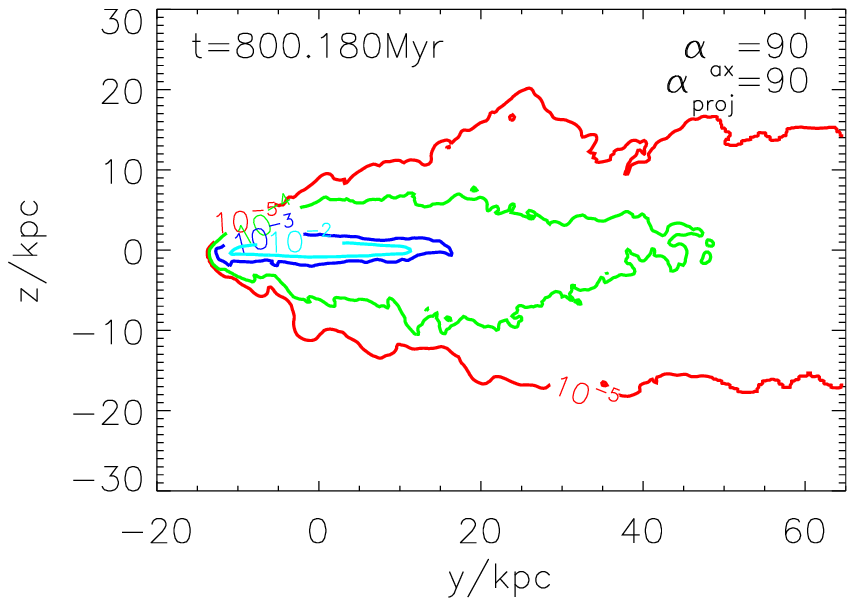}
\includegraphics[width=0.32\textwidth]{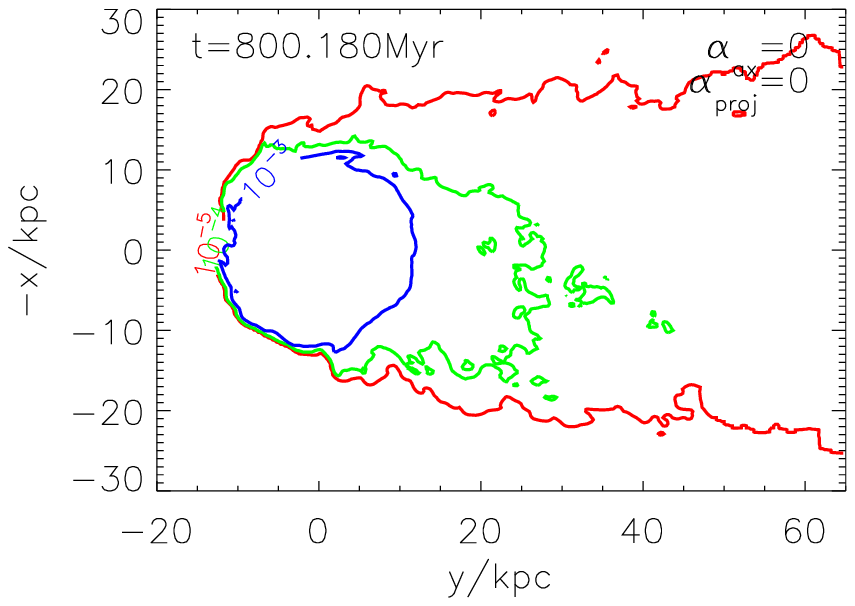}
\includegraphics[width=0.32\textwidth]{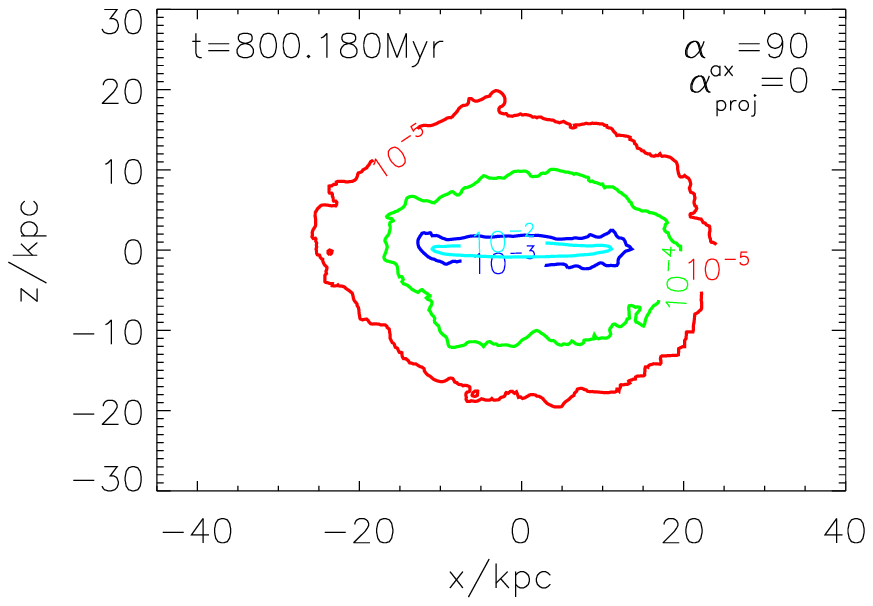}
\caption{Projected gas densities along $x$-, $z$-, and $y$-axis. For medium
  ram pressure and inclination of $90\degree$. See also
  Fig.~\ref{fig:proj_dens_i30}.}
\label{fig:proj_dens_i90}
\end{figure*}

\begin{figure*}
\includegraphics[width=0.32\textwidth]{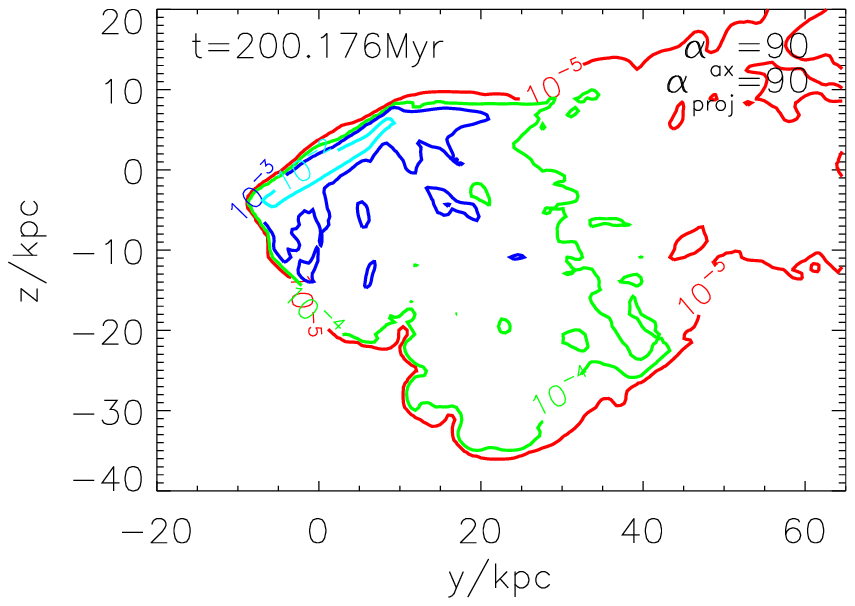}
\includegraphics[width=0.32\textwidth]{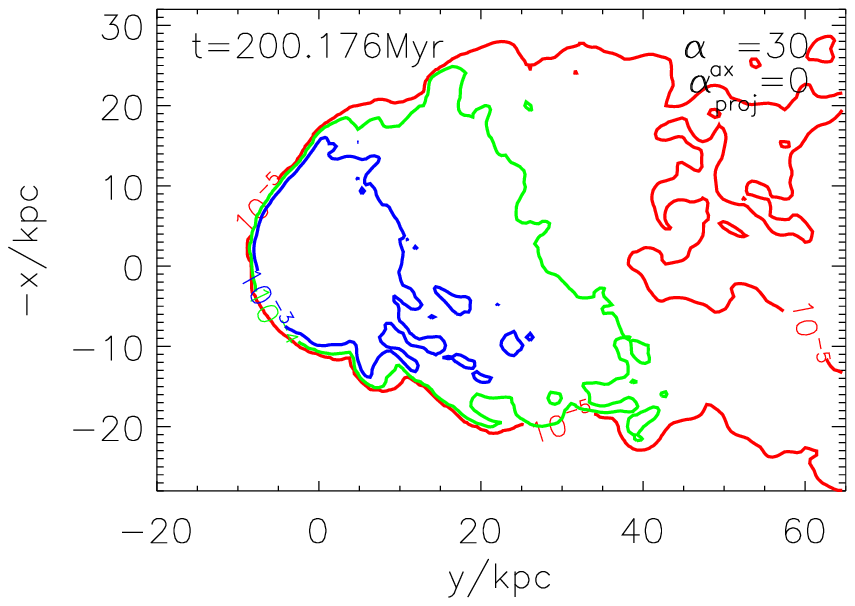}
\includegraphics[width=0.32\textwidth]{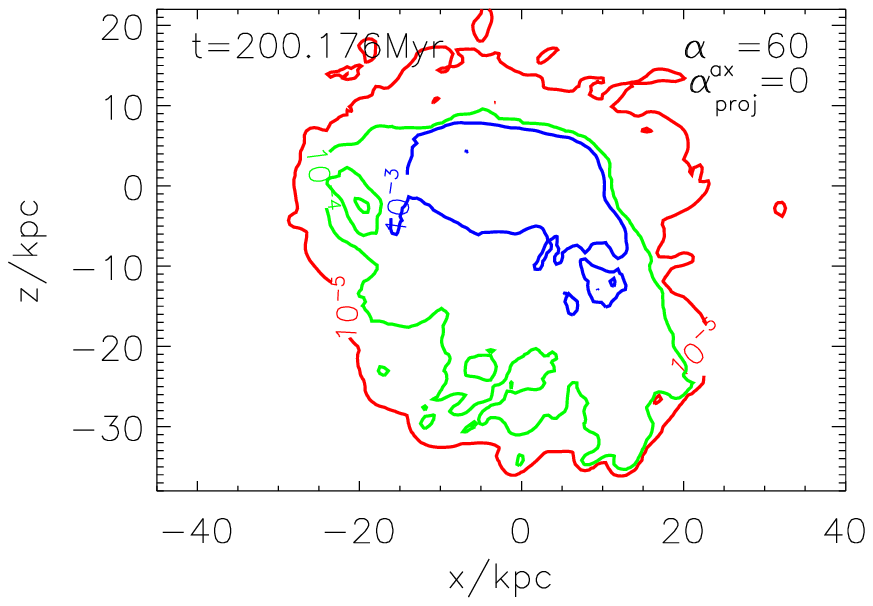}
\caption{Projected gas densities along $x$-, $z$-, and $y$-axis. For medium
  ram pressure and inclination of $60\degree$. See also
  Fig.~\ref{fig:proj_dens_i30}.}
\label{fig:proj_dens_i60}
\end{figure*}

In Figs.~\ref{fig:proj_dens_i30}, \ref{fig:proj_dens_i90} and
\ref{fig:proj_dens_i60}, we show projected gas densities at several timesteps
for the same runs as in Figs.~\ref{fig:slice_i30} and \ref{fig:slice_i90}, but
also for the corresponding $60\degree$ run. 
%
%
%
%
For all three runs, we show projections along the three coordinate axes. Only
galactic gas (identified by its ``colour'', see Sect.~\ref{sec:code}) is
shown.

The snapshots reveal an interesting trend: At early times a higher inclination
causes stronger asymmetry: one side (roughly the upstream side) is stripped
more strongly than the other. Details of how the rotation of the disc affects
the stripping are discussed below. However, even for the edge-on case, the
asymmetry decreases with time and the remaining gas disc becomes circular
again. This is caused by the rotation of the gas disc. At a galactic radius of
$10\Kpc$, the rotation time is roughly $300\Myr$, so in the course of our
simulation the disc rotates roughly 3 times. Therefore, for the case of a
constant ICM wind, eventually each part of the disc passes the region of
strongest stripping, and the disc becomes symmetrical again.

\subsection{Velocity information}
%
\citet{vollmer01a} stressed that velocity information is crucial for the
correct interpretation of HI observations of ram pressure stripped
galaxies. Due to the fact that we have used a constant ICM wind, the velocity
structure of the stripped material reflects mainly the
superposition of the galactic rotation and the acceleration by the ICM wind.
%
The velocity structure will be more interesting for simulations with
variable ICM winds.

\subsection{Radial profiles}  \label{sec:results_radial_profiles}
\begin{figure}
\centering\resizebox{0.75\hsize}{!}{\includegraphics{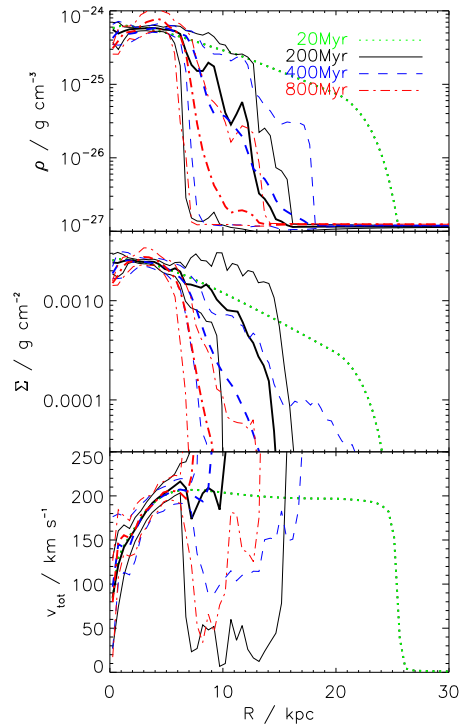}}
\caption{Evolution of radial profiles, for density, $\rho$, and rotation
  velocity, $v\Rot$, in the galactic plane. Also the projected gas surface
  density, $\Sigma$, is shown. The profiles belong to the same run as in
  Fig.~\ref{fig:slice_i30}. The thick line displays the mean profile averaged
  azimuthally over the disc, the thin lines display the minimum and maximum
  values for each radius. For further explanation see text,
  Sect.~\ref{sec:results_radial_profiles}. The colours/linestyles code four
  different times, see legend.}
\label{fig:R_profiles_evol}
\end{figure}

%
In Fig.~\ref{fig:R_profiles_evol}, we display the evolution of radial
profiles. 
%
%
To generate this plot, we measured radial profiles along 12 radial lines that
are separated by $30\degree$ in the galactic disc. The thick lines show the
average of these profiles, the thin lines maximum and minimum value for each
radius. The profiles reveal that the inner part of the gas disc remains
symmetrical, whereas the outer part becomes asymmetrical.

The decrease of the central density is caused by the spatial discretisation of
our grid. In grid-codes, advection of mass and momentum into non-axial
directions is always accompanied by a numerical viscosity due to numerical
diffusion. In the rotating gas disc, the gas has to move on circular orbits in
a Cartesian grid. This geometrical mismatch is strongest in the inner part of
the galaxy. Below a certain radius, the resolution of the rotational motion
becomes insufficient and the numerical diffusion introduces radial velocities,
which lead to a decrease of the central density. We checked in
higher-resolution runs that this behaviour does not influence our results (see
Appendix~\ref{sec:resolution}).

\subsection{Mass loss / radius decrease}

\subsubsection{Radius and mass measurement} \label{sec:radius_mass_measurement}
In the light of the previous plots, it is not clear how the radius of the gas
disc should be defined. We decided to include the asymmetry in our
measurement. We scan along 12 radial lines, separated by an angle of
$30\degree$, in the galactic plane. For each of these scan lines, we find the
radius where the density drops below $10^{-26}\gccm$ for the first time. This
is the stripping radius for this scan line. This density limit seems rather
arbitrary, but \citet{mythesis} tested different options and found this one to
be the most appropriate and representative one. We average the results of all
scan lines to obtain a mean disc radius $R\Mean$, but we also compute the
maximum and minimum radius $R\Max$ and $R\Min$.

Two masses are of interest here: the mass $M\Disc$ of gas inside a cylinder
centred on the galaxy (radius=$27\Kpc$, thickness = $\pm 5\Kpc$), and the mass
of gas bound to the galaxy, $M\Bound$. In order to exclude the ICM in
$M\Disc$, we only consider gas with a temperature less than $10^7\K$. The
bound mass is given by the sum of the gas mass in all those grid cells where
the total energy density $e\Therm + e\Kin + e\Pot = p/(\gamma-1) + 0.5 \rho
(v_x^2 + v_y^2 + v_z^2) + \rho\Phi$ is negative.

\subsubsection{Radius and mass}
\begin{figure*}
\includegraphics[width=0.32\textwidth]{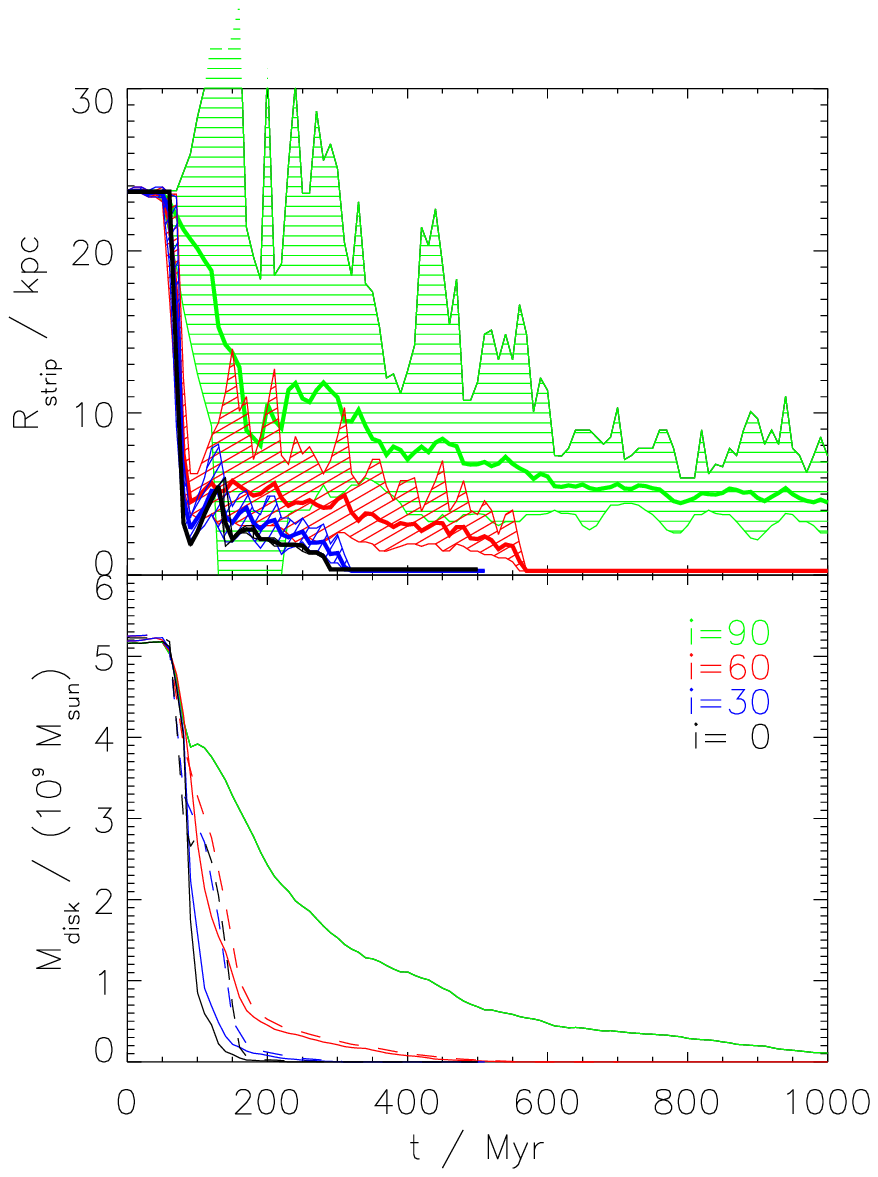}
\includegraphics[width=0.32\textwidth]{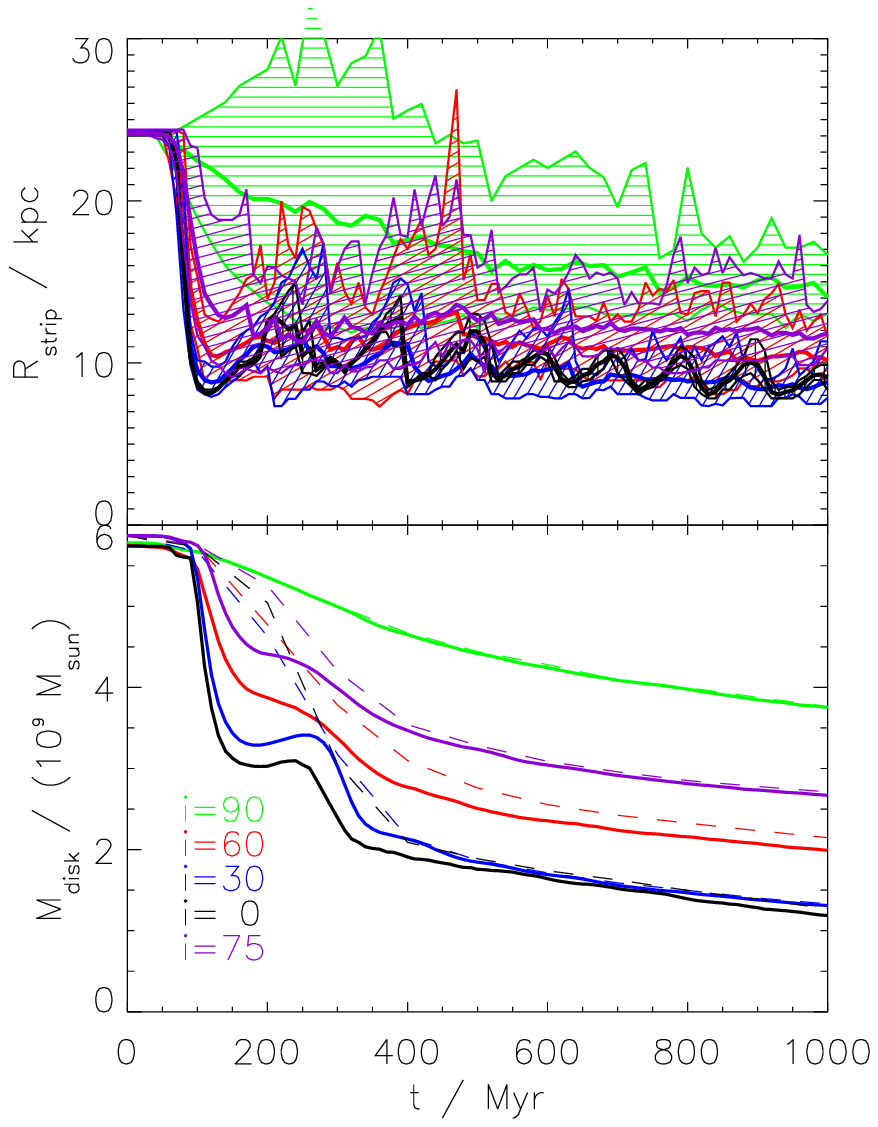}
\includegraphics[width=0.32\textwidth]{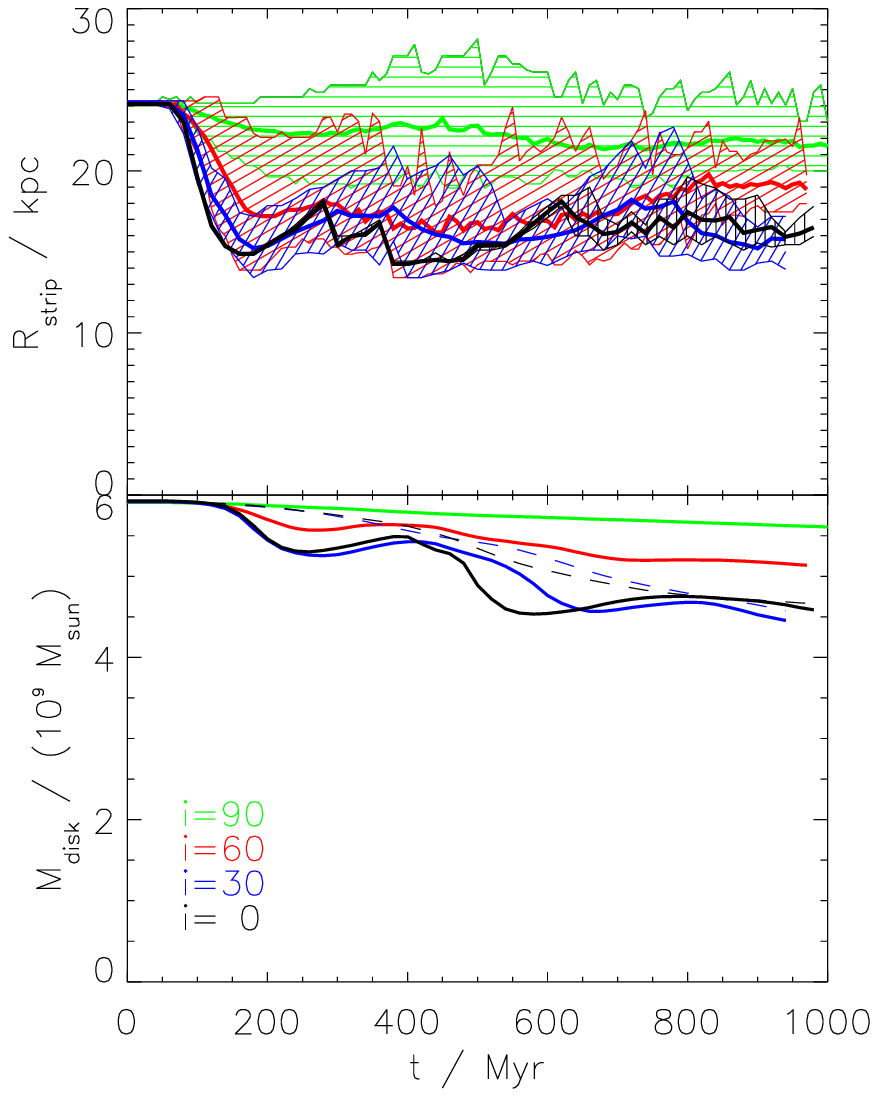}
\caption{Radius and mass of the remaining gas disc for three ram pressures
  (left, middle and right plot correspond to strong, medium and weak ram
  pressure) and different inclinations (colour-coded). For the radius, we show
  the mean radius (thick solid line) and the minimum and maximum radius (thin
  lines of same colour). The area between the minimum and maximum radius
  curves is hatched for easier orientation. In addition to the colour, also
  the direction of hatching codes the inclination (e.g. horizontal hatching
  for edge-on). Thin dashed lines in mass plot are bound gas mass, thick solid
  lines are the mass in a fixed disc region (see text,
  Sect.~\ref{sec:radius_mass_measurement}). Both, mass and radius do not drop
  immediately because we switch on the flow at the inflow boundary at $t=0$,
  and then the flow needs a certain time to reach the galaxy.}
\label{fig:radius_mass_evol}
\end{figure*}

\begin{figure*}
\includegraphics[width=0.32\textwidth]{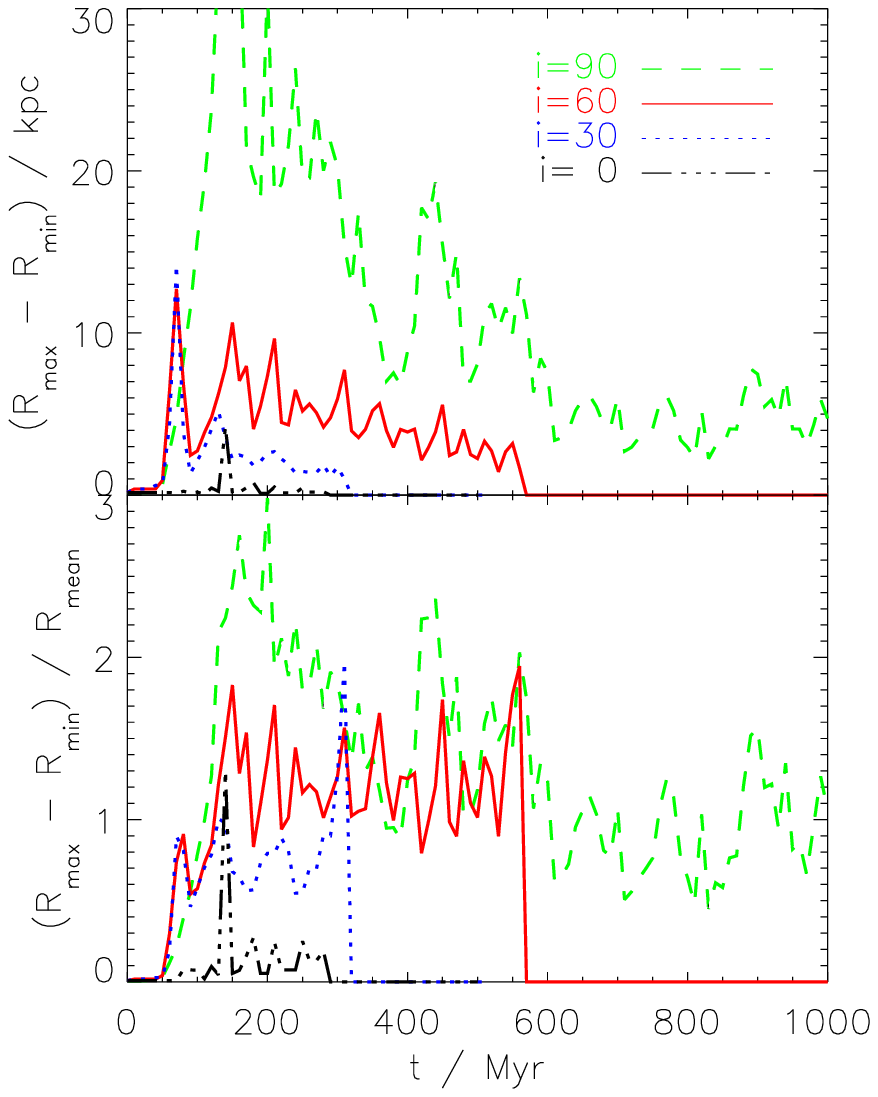}
\includegraphics[width=0.32\textwidth]{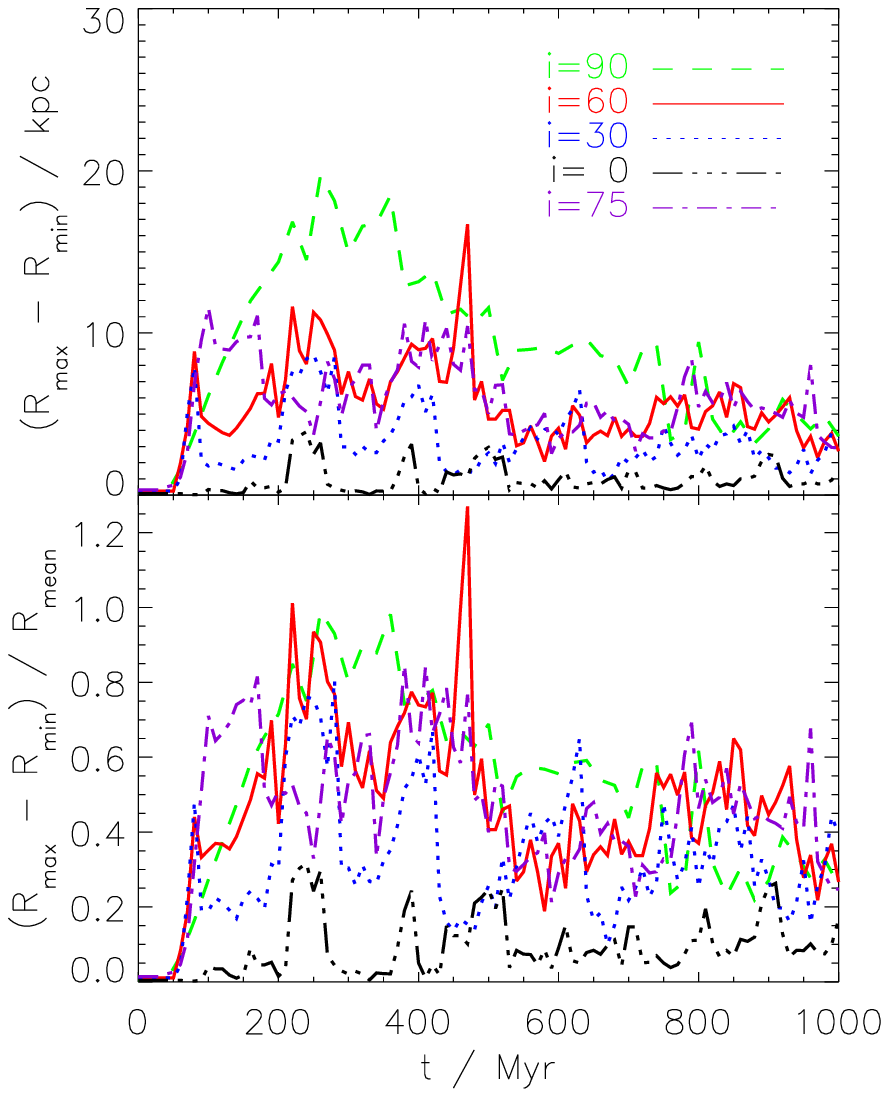}
\includegraphics[width=0.32\textwidth]{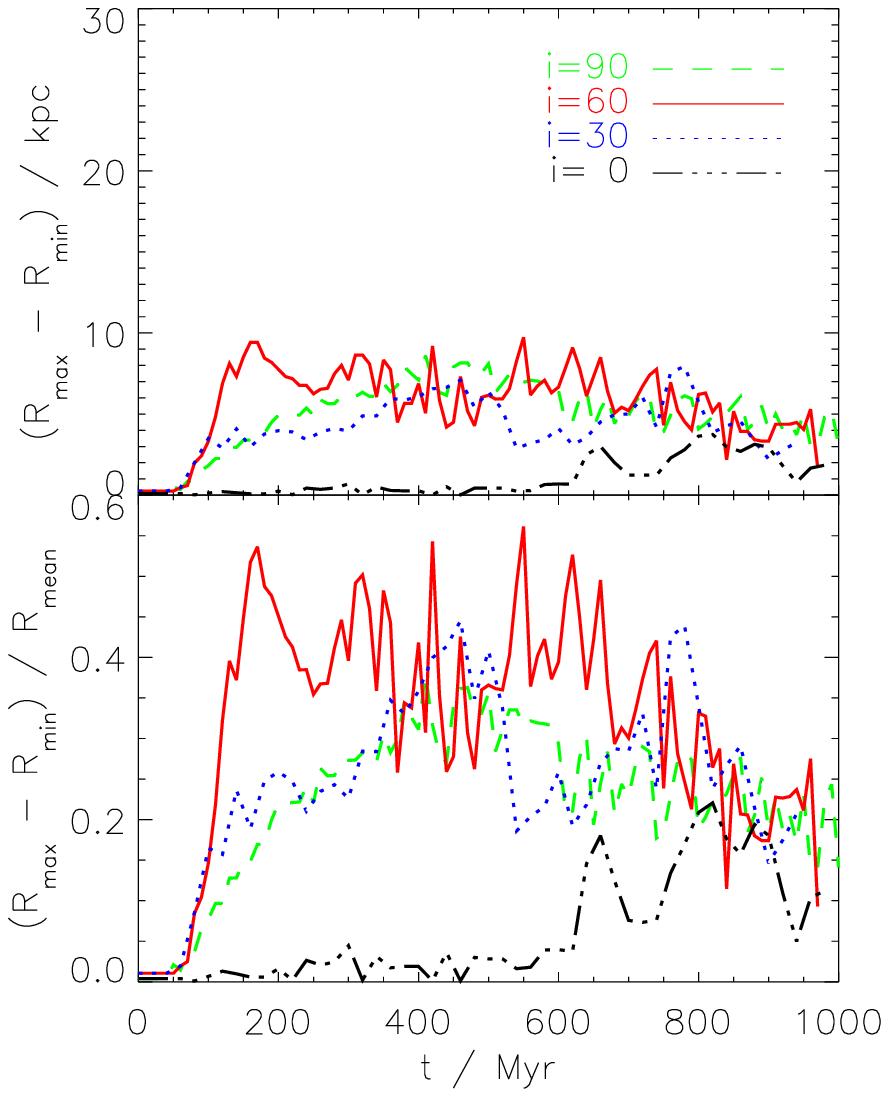}
\caption{Demonstration of the asymmetry of the gas disc and its evolution. The
  asymmetry is shown by $R\Max - R\Min$ in the top panels, and the same
  normalised to $R\Mean$ in the bottom panels. For three ram pressures: left,
  middle and right plot correspond to strong, medium and weak ram
  pressure. In each plot, the results for different inclinations are shown
  (coded by colour/linestyle, see legend). Please note the different scales on
  the $y$-axes of the bottom panels.}
\label{fig:asymmetry}
\end{figure*}

The evolution of disc mass and radius is shown in
Fig.~\ref{fig:radius_mass_evol}. 
%
%
The left, middle and right plot correspond to strong, medium and weak ram
pressure, the inclinations are colour-coded. The reason why the initial mass
appears less in the case of the high ram pressure is that we have set the gas
disc in pressure equilibrium with the ICM. For a higher $\rho\ICM$ but
constant ICM temperature, the ICM pressure is higher, which causes the outer
layers of the gas disc to have a higher pressure. As the density distribution
is fixed, this means the outer layers have a higher temperature and, therefore,
drop out of our calculation of $M\Disc$ due to the temperature limit (see also
discussion of this effect in \citealt{roediger05}).  In order to include the
asymmetry of the disc, we plot $R\Min(t)$ and $R\Max(t)$ as a function of
time. In addition, we overplot the mean radius, $R\Mean(t)$. For strong and
medium ram pressure, it is obvious that the degree of asymmetry increases with
inclination. We have also plotted the quantities $R\Max - R\Min$ and $(R\Max -
R\Min)/R\Mean$ in Fig.~\ref{fig:asymmetry} to quantify the degree of
asymmetry.
%

Several interesting statements can be made from the radius and mass plots:
\begin{itemize}
\item Depending on ram pressure, we can strip the disc completely, partially or
  only marginally. 
\item The evolution of the retained mass and, particularly, the disc radius
  are remarkably similar for face-on, $30\degree$ and $60\degree$ runs. The
  runs for $0\degree$ and $30\degree$ are nearly indistinguishable, and in the
  $60\degree$ case only slightly less mass is lost. 
\item For all but the edge-on runs, we can clearly distinguish the three phases
  discussed in \citet{roediger05} and in Sect.~\ref{sec:intro}: \newline
  The first is the instantaneous stripping phase, during which the outer part
  of the gas disc is pushed in the downwind direction. This phase ends with the
  first minimum in $R\Mean(t)$. The first minimum in $M\Disc(t)$ appears
  slightly later because the gas has to leave our defined ``disc
  region'' first before it is no longer considered as disc gas.\newline
  The next phase is the intermediate phase, which is characterised by
  $M\Bound(t)>M\Disc$.  \newline
  Finally, the phase of continuous stripping follows.
\item The durations of, both, the instantaneous stripping phase and the
  intermediate phase are longer for weaker ram pressures. For a fixed
  $p\Ram$ they are independent of inclination.
\item For non-edge-on cases, the evolution of $R\Mean$ {\em after} the
  instantaneous phase is slow.
\item For a given $p\Ram$, the mass loss rate in the continuous stripping phase
  seems to be independent of inclination. This feature is most relevant for
  medium ram pressures, where neither the mass loss is negligible (like for
  weak ram pressures) nor the disc is stripped completely (like for the strong
  ram pressure).
\end{itemize}
The basic point in all these aspects is that the stripping process is nearly
independent of inclination as long as the inclination is not close to
edge-on. 

The dependence of asymmetry on ram pressure and inclination is more
complex. Moreover, the asymmetry decreases with time. In general, the face-on
cases show nearly no asymmetry. The absolute asymmetry (i.e. $R\Max - R\Min$)
for the edge-on cases depends strongly on ram pressure, where higher ram
pressures lead to significantly higher asymmetries. For the remaining
inclinations, the absolute asymmetry is approximately independent of ram
pressure. The dependence of the relative asymmetry (i.e. $(R\Max -
R\Min)/R\Mean$) on ram pressure reflects the dependence of $R\Mean$ on
$p\Ram$. The radial profiles in Fig.~\ref{fig:R_profiles_evol} reveal that the
asymmetry is restricted to the outer part of the gas disc. The inner part
remains symmetrical. This is even true for the medium ram pressure edge-on
case, as can be seen in Fig.~\ref{fig:slice_i90}. Even for inclined cases, an
inner part of the gas disc remains more or less untouched.

The comparison between subsonic and supersonic runs with identical ram
pressure reveals a similar result to the result of \citet{roediger05}. In the
supersonic case, the galaxy can retain slightly more mass than in the subsonic
case. The difference in retained masses increases with inclination. For highly
inclined galaxies, also the asymmetry of the disc is temporarily larger in the
supersonic case.

\subsection{Comparison analytical/numerical} \label{sec:comp_analyt_num}
%
Due to the asymmetry of the disc and the ongoing evolution, it is not
straightforward to compare the numerical results to the analytical
predictions. A first question to answer is at which time we have to measure
mass and radius in our simulations to make a meaningful comparison to the
analytical prediction. We take measurements at two timesteps: at the end of
the instantaneous phase and at the end of the intermediate phase. For the
strong ram pressure, the intermediate phase is finished at $t=175\Myr$, for the
medium ram pressure at $t=400\Myr$, and for the weak one at
$t=800\Myr$. Except for the edge-on cases, the stripping radius does not
change during the intermediate phase. Therefore, the radius measurements taken
at the end of of the intermediate phase are also representative for the end of
the instantaneous phase. The disc mass, however, changes during the
intermediate phase. As a further complication, $R\Max(t)$ and $R\Min(t)$
oscillate on rather short timescales. We average these quantities over a
smoothing length of $100\Myr$ to derive typical values.

\begin{figure}
\centering\resizebox{0.75\hsize}{!}{\includegraphics[angle=0]{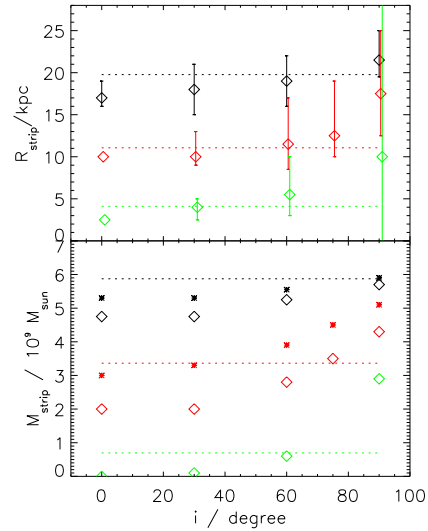}}
\caption{The analytical estimate of stripping radius and stripping mass as a
  function of inclination for three different ram pressures (black=weak,
  red=medium, green=strong) is shown by dotted lines. Also the numerical
  results are shown. We measure the numerical stripping radius and mass at the
  end of the intermediate phase, see text for a more detailed discussion. For
  the numerical radius we give the mean radius as a diamond, and the range
  between the minimum and maximum measured radius as an error bar. See
  text for further explanation (Sect.~\ref{sec:comp_analyt_num}). The disc
  mass at the end of the intermediate phase is shown by diamonds. In
  addition, we plot the disc mass at the end of the instantaneous phase as
  stars.}
\label{fig:analytic}
\end{figure}

In Fig.~\ref{fig:analytic}, we show the numerical results and the analytical
estimate for the stripping radius and the remaining mass in the
disc. 
%
%
The analytical estimate is displayed by the dotted lines. The predicted
stripping mass given in the lower panel of this plot is simply the disc mass
inside the stripping radius. For the numerical stripping radius, we show
$R\Mean$ as diamonds and the range $R\Min$ to $R\Max$ as an error bar. For the
mass remaining in the disc, we show the numerical results at the end of the
intermediate phase as diamonds. The disc mass at the end of the
instantaneous phase is shown by stars. However, for the high ram pressure case
this quantity cannot be derived.

For the face-on cases ($i=0$), the analytical estimate and the numerical
result for the stripping radius agree well. The same is true for the disc mass
at the end of the instantaneous phase. The numerical values are just slightly
smaller than the predicted ones. Here we confirm the result of previous
works. At the end of the intermediate phase the galaxy has already lost more
than the gas outside $R\Strip$. Therefore, the numerical values for the disc
mass at the end of the intermediate phase are significantly lower than the
analytical prediction.

For inclinations $\lesssim 30\degree$, neither stripping radius nor disc mass
depend on inclination. For $30\degree < i \lesssim 60\degree$ the dependence
on inclination is weak. Only for $i\gtrsim 60\degree$ the stripping radius and
disc mass increase with inclination. Therefore, the simulation results confirm
the prediction made in Sect.~\ref{sec:analytic_inclination}.

The mass loss rate in the continuous stripping phase is remarkably similar in
all cases and is about $1M\Sun/\Yr$. This is similar to the results of
\citet{roediger05}.

\section{Discussion}
\label{sec:discussion}
%
\subsection{The influence of the inclination}
%
According to our simulations, the inclination angle does not affect stripping
significantly unless the galaxy moves close to edge-on. This result agrees with the
trends found by \citet{quilis00} and \citet{marcolini03}. In contrast to
\citet{quilis00}, we do not find that only strict edge-on galaxies suffer less
stripping, but we find that $i$ affects the amount of mass loss for
inclinations within $\sim 30\degree$ of the edge-on geometry. Moreover, we
have studied the stripping process for longer runtimes and investigated a
larger range of ram pressures.
 
Our result disagrees with \citet{schulz01} who find that inclined galaxies are
stripped on a longer timescale, also for inclinations around $45\degree$. In
contrast, in our simulations the phases of the ICM-ISM interaction are
remarkably independent of inclination. The duration of the phases is
comparable to what was found by \citet{roediger05}.

\citet{vollmer01a} have included a time-dependent ram pressure in their
simulations, which seems to make the influence of inclination rather
messy. They could not find a correlation between ram pressure, inclination
angle and stripping radius (or remaining disc diameter), but they give an
empirical fit for the remaining mass. We checked whether our simulation
results for the disc masses follow their suggestion, but this is not the case.

Based on their study of dwarf galaxies, \citet{marcolini03} argue that the
inclination is only important when the ram pressure is comparable to the
central pressure $p_0$ of the galactic potential well. For $p\Ram > p_0$ the
galaxy will be stripped completely, independent of inclination, and for $p\Ram
< p_0$ only very little gas is lost.  For our galaxy, the central pressure is
$p_0 \approx 10^{-10}\Erg\ccm$. Our ram pressures are $6.4\cdot
10^{-11,-12,-13}\Erg\ccm$. If only the central galactic pressure is compared
with $p\Ram$, a dependence on inclination should be prominent only in the case
of our strongest ram pressure. In our simulations, however, at least
the strong and the medium ram pressure case clearly depend on inclination. We
suggest that not only the central disc pressure is relevant for this
comparison, but the pressure profile along the galactic plane. For our model
galaxy, the pressure along the galactic plane drops to about $10^{-11}\Erg\ccm$
at a radius of $5\Kpc$ and down to about $10^{-12}\Erg\ccm$ at a radius of
$10\Kpc$. This explains why the inclination plays a role for the strong and
the medium ram pressure. For the weak ram pressure, our galaxy loses little
mass for all inclinations. However, even there, the influence of the
inclination is the same.  We conclude that the result of \citet{marcolini03}
cannot be applied to massive galaxies directly but needs to be improved in the
sense that the inclination is important as long as the ram pressure is
comparable to the pressure in the galactic plane.

\subsection{What could be observed?}
%
A prediction for observable gas surface density maps is given in
Figs.~\ref{fig:proj_dens_i30} to \ref{fig:proj_dens_i60}. We want to
emphasise one special aspect, namely the relation between the direction of the
tail of stripped gas and the galaxy's direction of motion. Consider, e.g., the
following surface density maps (all at $t=200\Myr$):
Fig.~\ref{fig:proj_dens_i30}, first line and first column;
Fig.~\ref{fig:proj_dens_i90} second line and second column; and
Fig.~\ref{fig:proj_dens_i60} first colum. Also focus on contour lines of
$10^{-4}\,\mathrm{g}\Cm^{-2}$ and above, as it would be the case for
``normal'' HI observations. Now try to guess the galaxy's direction of motion
purely from these maps, from the direction of the tail. The answer is {\em
not} that the galaxy is moving into the direction opposite to the tail, but in
all cases the galaxy is moving towards the left.  The reason for this strange
appearance is simple for the two cases where the disc is seen edge-on: Going
back to the rest frame of the galaxy, the ICM wind is blowing the stripped gas
towards the right. The gas from the upwind edge has to go a longer way to
reach a certain distance to the galaxy in $y$-direction, whereas the gas from
the downwind edge has a ``head start''. Moreover, in the direct vicinity of
the galaxy, the ICM wind is not blowing straight in $y$-direction, but it is
flowing smoothly past the galaxy. This introduces velocity components in $\pm
x$ and $\pm z$ direction. Therefore, for some time during the initial phase
the tail of stripped gas appears to be in the ``wrong'' direction. For such
cases, not even radial velocity information would help to determine the true
direction of motion because this behaviour is governed completely by
velocities perpendicular to the line-of-sight.

In the case of the map in Fig.~\ref{fig:proj_dens_i90}, second line and second
column, the reason is a different one. This galaxy is moving edge-on through
the ICM, while we see it face-on. If we can determine the position of the
galactic centre, e.g. from the stellar disc, we would see that the gas disc is
stretched towards the top-right and compressed at the opposite side. But also
this galaxy is moving towards the left. In this example, we can see the
interaction between the galactic rotation and the ICM wind. In this map, the
galaxy is rotating counter-clockwise. The side where the gas rotates parallel
to the ICM wind is stripped more easily than the side where the gas rotates
antiparallel to the wind. In later stages (further maps in the
middle column of Fig.~\ref{fig:proj_dens_i90}), we can see that the tail of
stripped gas seems to be attached to the bottom side of the
galaxy. \citet{phookun95} discussed the effects of combined stripping and
rotation and concluded that in such a geometry the side where the gas rotates
{\em into} the wind will be stripped more strongly because, as the relative
velocity between the gas and the ICM is higher at that side, also the true ram
pressure is higher at this side. Thus, the tail of stripped gas should be
twisted towards this more strongly stripped side. This is just the opposite of
what our simulations show. The relative velocity is higher at the side where
the gas rotates into the wind, and also the ram pressure may be stronger
there. But at this side the gas first has to be decelerated to zero velocity
and then accelerated to get stripped. At the opposite side, the gas is already
moving in wind direction and, hence, it seems easier to strip it there.

In all cases discussed here, deep  HI observations that reveal the low
density extension of the tail or, for supersonic cases, the observation of a
bow shock could help to determine the true direction of motion.

\subsection{Application to cluster galaxies}
%
The strongest limitation of our approach is the use of a constant ICM wind,
which is not a good model for galaxies that traverse the cluster
centre. Nonetheless, some of our results are applicable to real cluster
galaxies.

According to our simulations, the mass loss is similar for galaxies that do
not move close to edge-on also in initial phase of stripping. Therefore, even
if galaxies are exposed to strong ram pressures for a short period olny, only
those galaxies that are close to edge-on can be protected against stripping by
their inclination. Most galaxies should suffer severe stripping in cluster
centres.

Although the mass loss from galaxies in low ram pressure environments such as
cluster outskirts is small, \citet{roediger05} found that still the gas discs
are truncated at $\sim 15$ to $20\Kpc$. While this work was restricted to
face-on geometries, we found that this is also true for all
non-close-edge-on geometries. Thus, ram pressure stripping of massive galaxies
may also play a role in cluster outskirts. Observational evidence for the
``pre-processing'' of galaxies in low-density environments (see also
\citealt{fujita04}) arises from the detection of HI deficient galaxies
(\citealt{solanes01}) and passive spirals (\citealt{goto03a}) in cluster
outskirts.

The duration of the intermediate phase, which is the phase where a substantial
amount of gas that will be stripped eventually, is still bound to the
galaxy, is independent of inclination. Therefore, we expect that if the ICM
wind strength decreases during this phase, as it could be the case for
galaxies passing cluster centres, a substantial amount of the stripped gas will
still be bound to the galaxy and will fall back to the disc. This effect was
found by \citet{vollmer01a} in their sticky particle simulations. They also
found that the amount of gas that falls back into the galaxy increases with
inclination. It will be interesting to compare this feature with
hydrodynamical simulations of a cluster passage.

If the ICM wind decreases during the intermediate phase, it leaves behind a gas
disc with a strong asymmetry in the outer parts. Differential rotation should
smear out this asymmetry on a few galactic rotation timescales, resulting in a
gas disc with a steepened gas density profile in the outer parts. A similar
feature has been observed recently by \citet{koopmann05} in the H$\alpha$
discs of Virgo spirals. We suggest that this feature could be caused by the
scenario we just described.

\section{Summary}
%
We have presented 3D hydrodynamical simulations of ram pressure stripping of
massive disc galaxies. We concentrated on the influence of the inclination
angle between the galactic rotation axis and the galaxy's direction of motion.

We find that the inclination angle has a weak effect on the mass loss from the
gas disc as long as the galaxy is not moving close to edge-on. We can explain
this behaviour analytically by comparing the effective ram pressure with the
component of the gravitational restoring force in wind direction.  From
simulations of dwarf galaxies, \citet{marcolini03} concluded that the
inclination angle does not play a role as long as the ICM ram pressure
is not comparable to the central disc pressure, $p_0$. For $p\Ram < p_0$ hardly
any gas would be lost, and for $p\Ram > p_0$ the gas disc would be stripped
completely for all inclinations. We refine this result in the sense that for
massive galaxies it is not sufficient to compare the central disc pressure
with $p\Ram$, but the pressure profile along the galactic plane must be used
for this comparison. This leads to a larger range of ram pressures for which
the inclination angle plays a role.

We have also studied the phases of the ICM-ISM interaction and found that for
all non-edge-on geometries the stripping proceeds in a remarkably similar
fashion: First, during the instantaneous stripping phase, the outer part of the
gas disc is pushed towards the downstream direction. Then follows the
intermediate phase which is needed to unbind the stripped gas from the
galactic potential. Finally, during the continuous stripping phase the gas disc
continues to lose gas at a small rate of $\sim 1M\Sun \Yr^{-1}$.

In addition to the mass loss, we have studied the asymmetrical structure
introduced in the remaining gas discs of inclined galaxies.

We have also presented projected gas density maps for several simulation
runs. We have demonstrated that in moderately deep HI observations the tail of
stripped gas does not necessarily point in a direction opposite to the
galaxy's direction of motion. Therefore, the observation of a galaxy's gas
tail may be misleading about the galaxy's direction of motion. Deep HI
observations or additional observations of, e.g., a bow shock may be essential.


\section*{Acknowledgements}
We acknowledge the support by the DFG grant BR 2026/3 within the Priority
Programme ``Witnesses of Cosmic History'' and the supercomputing grants NIC
1927 and 1658 at the John-Neumann Institut at the Forschungszentrum J\"ulich. 
Some of the simulations were produced with STELLA, the LOFAR BlueGene/L System.

The results presented were produced using the FLASH code, a product of the DOE
ASC/Alliances-funded Center for Astrophysical Thermonuclear Flashes at the
University of Chicago. 

We gratefully acknowledge fruitful and helpful discussions with Joachim
K\"oppen and Gerhard Hensler.

\appendix
\section{Resolution comparison}
\label{sec:resolution}
\begin{figure*}
\includegraphics[trim=0 80 0 0,clip,angle=0,width=0.32\textwidth]{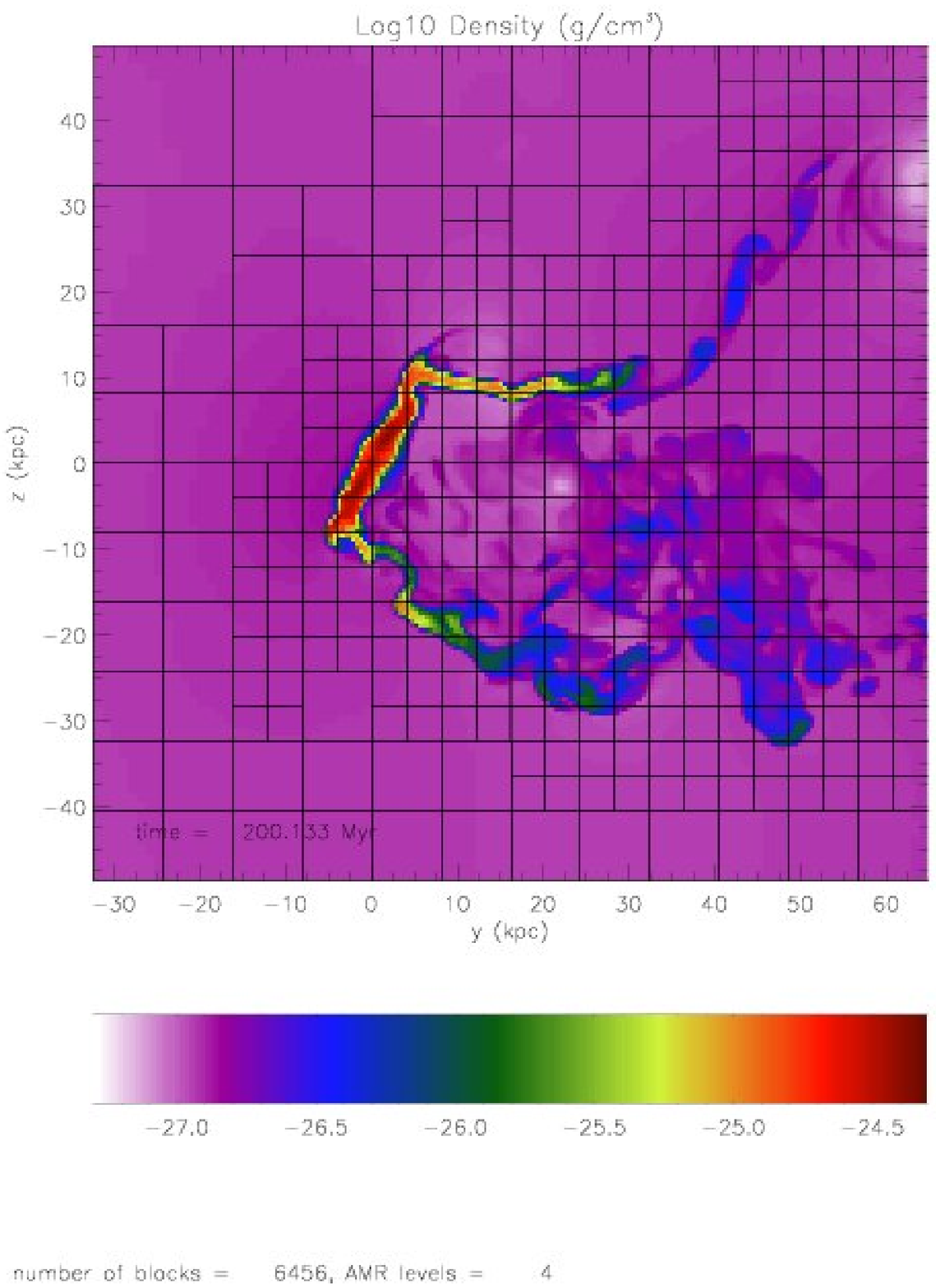}
\includegraphics[trim=0 80 0 0,clip,angle=0,width=0.32\textwidth]{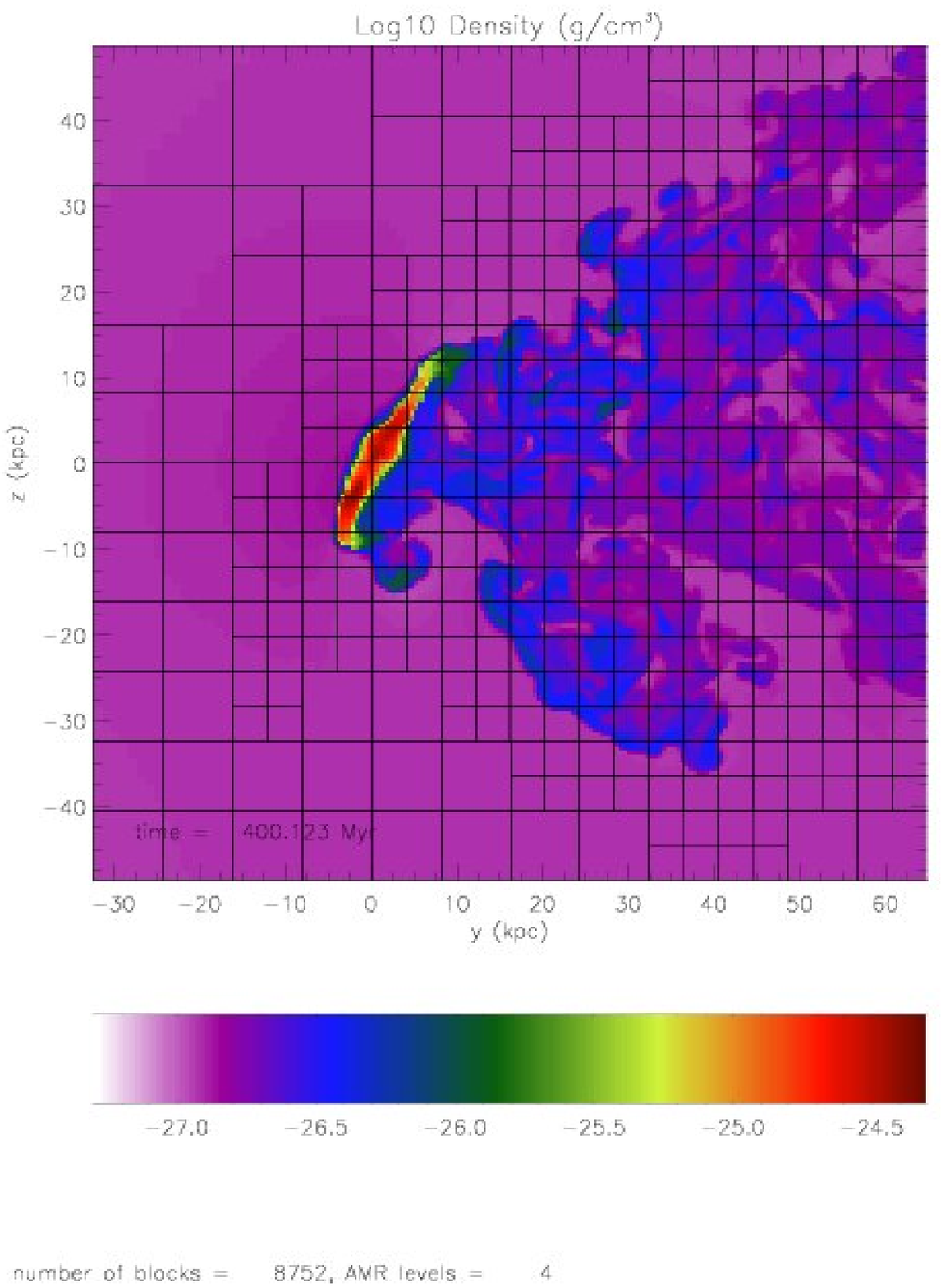}
\includegraphics[trim=0 80 0 0,clip,angle=0,width=0.32\textwidth]{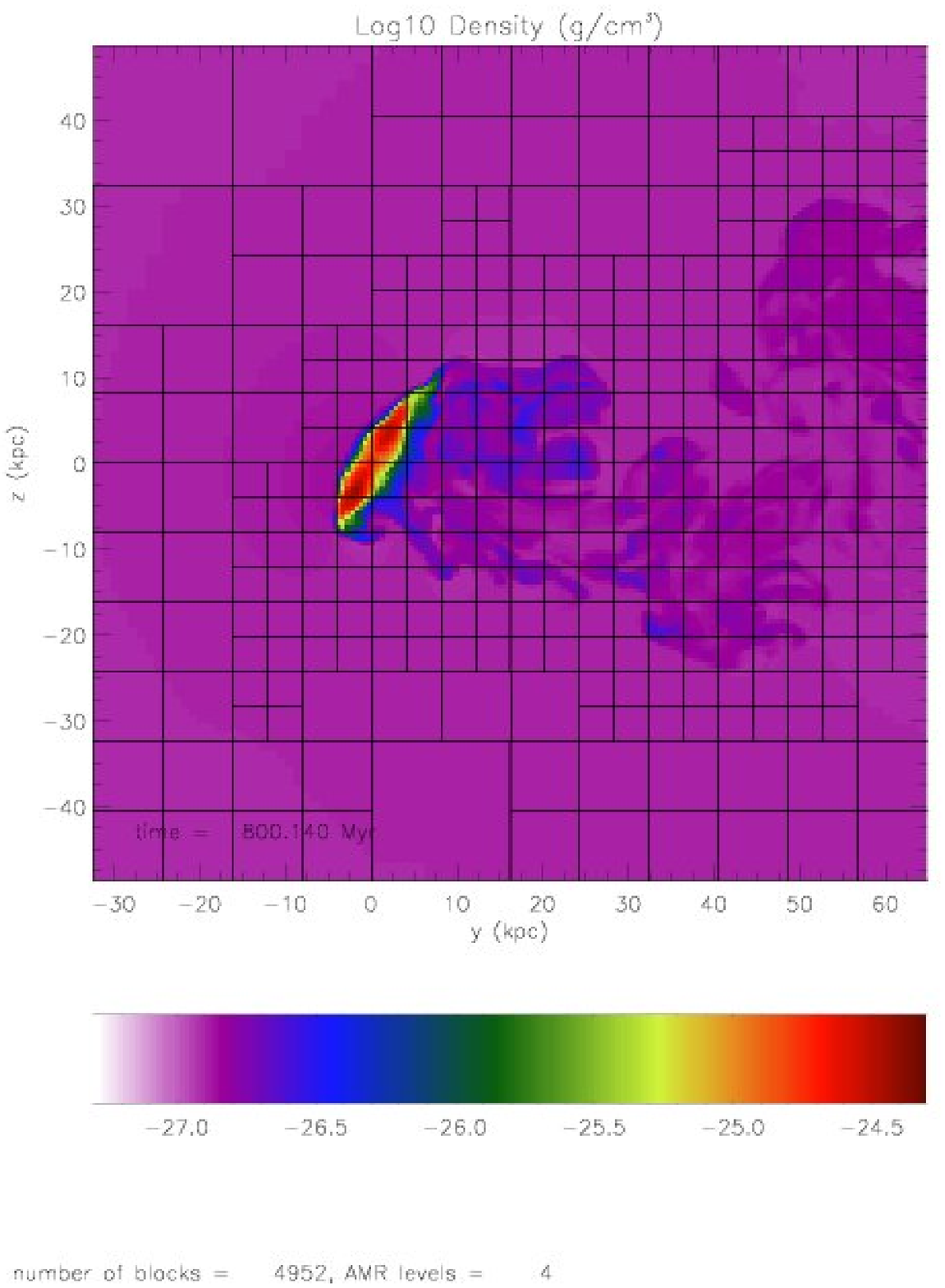}
\caption{Same as Fig.~\ref{fig:slice_i30}, but for lower resolution ($\Delta x
  = 500\PC$).}
\label{fig:slice_i30_LR}
\end{figure*}

\begin{figure*}
\includegraphics[trim=0 80 80 0,clip,angle=0,width=0.32\textwidth]{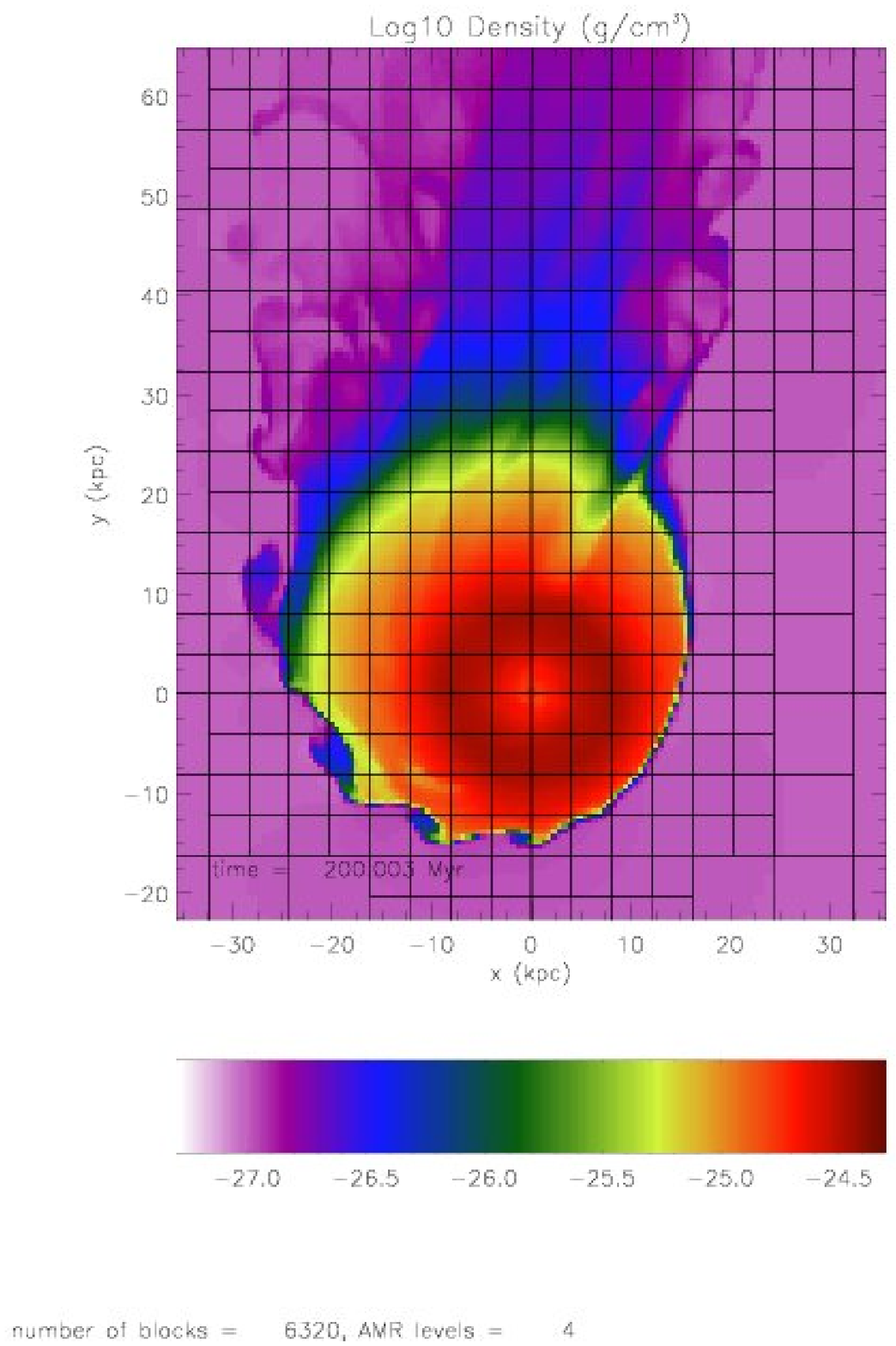}
\includegraphics[trim=0 80 80 0,clip,angle=0,width=0.32\textwidth]{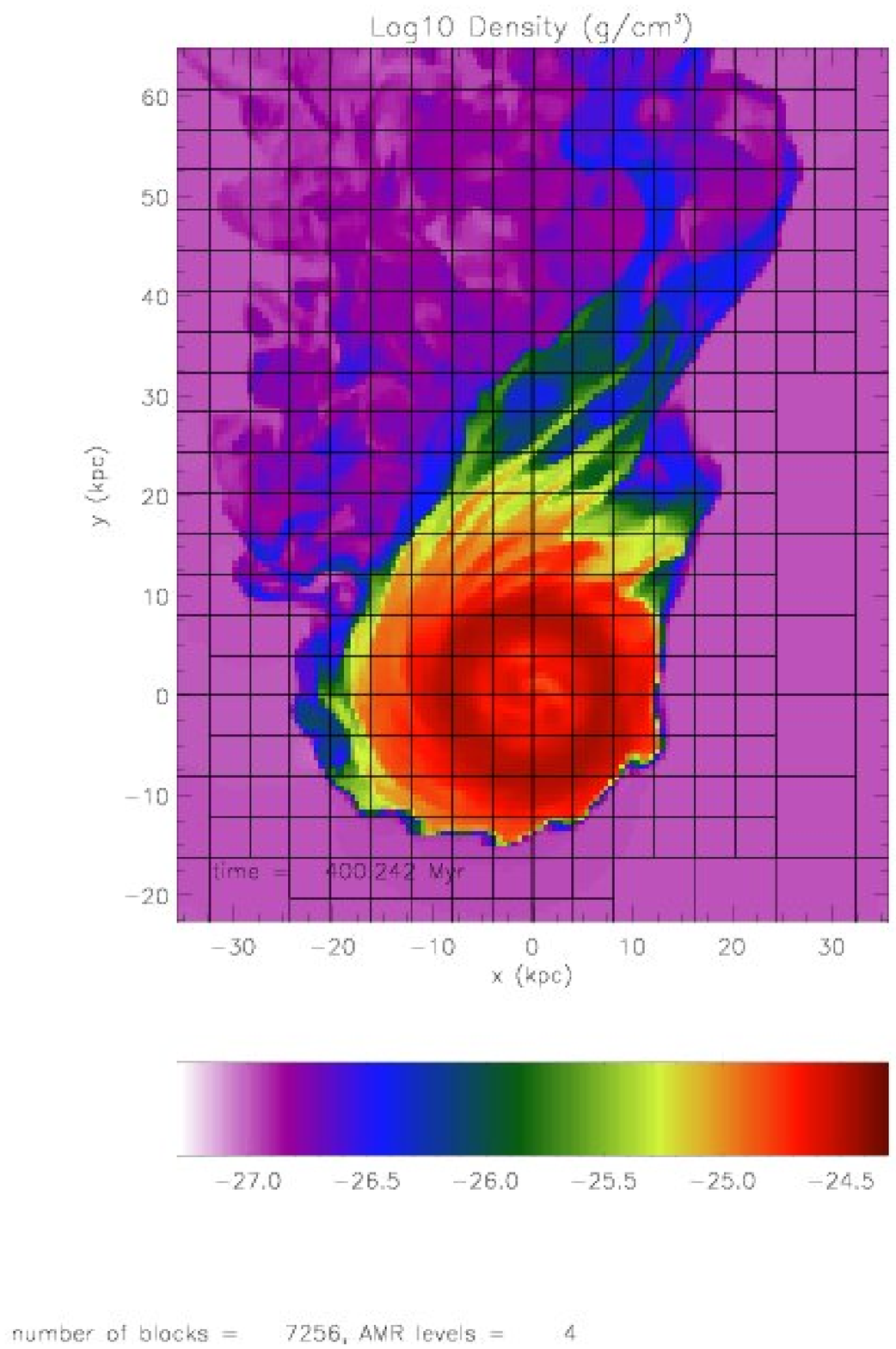}
\includegraphics[trim=0 80 80 0,clip,angle=0,width=0.32\textwidth]{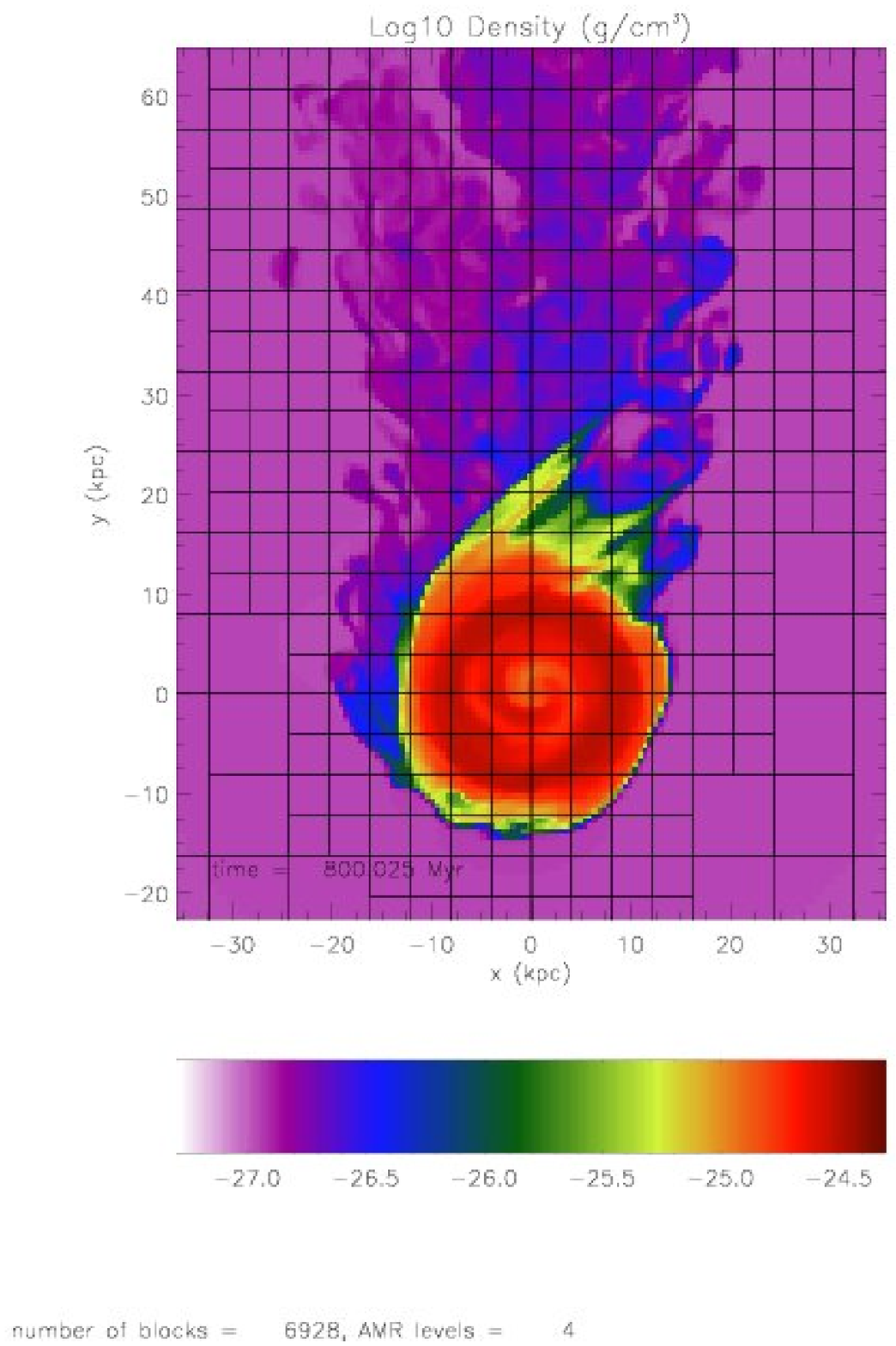}
\caption{Same as Fig.\ref{fig:slice_i90}, but for lower resolution run
  ($\Delta x = 500\PC$).}
\label{fig:slice_i90_LR}
\end{figure*}

We have performed some of our simulation runs with two different resolutions
to verify that our results are resolution independent. In
Figs.~\ref{fig:slice_i30_LR} and \ref{fig:slice_i90_LR}, we repeat slices
through the simulation box corresponding to Figs.~\ref{fig:slice_i30} and
\ref{fig:slice_i90}, but for the lower resolution of $500\PC$.
%
%
%
In the higher resolution runs, the stripped gas fragments more than in the low
resolution runs, but the overall structure is very similar for both
resolutions. 

\begin{figure}
\centering\resizebox{0.75\hsize}{!}{\includegraphics{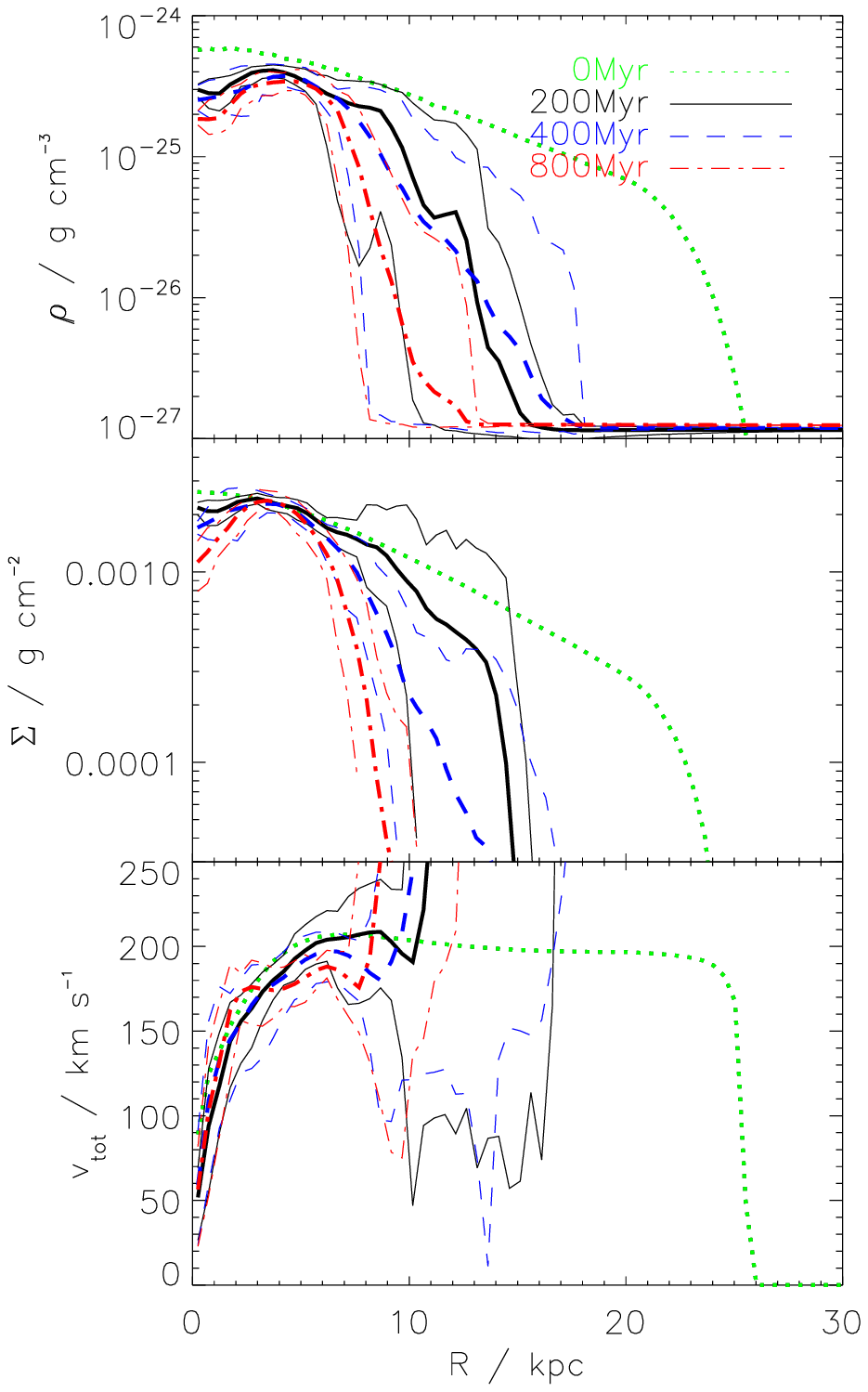}}
\caption{Same as Fig.~\ref{fig:R_profiles_evol}, but for low resolution run.}
\label{fig:R_profiles_evol_LR}
\end{figure}

In Fig.~\ref{fig:R_profiles_evol_LR}, we repeat Fig.~\ref{fig:R_profiles_evol}
but for the low resolution case. Here the numerical diffusion problem in the
inner part as explained in Sect.~\ref{sec:results_radial_profiles} occurs in a
larger region in the inner part. Aside from this, the characteristics of the
profiles are very similar for the two resolutions.
%

\begin{figure}
\centering\resizebox{0.75\hsize}{!}{\includegraphics[]{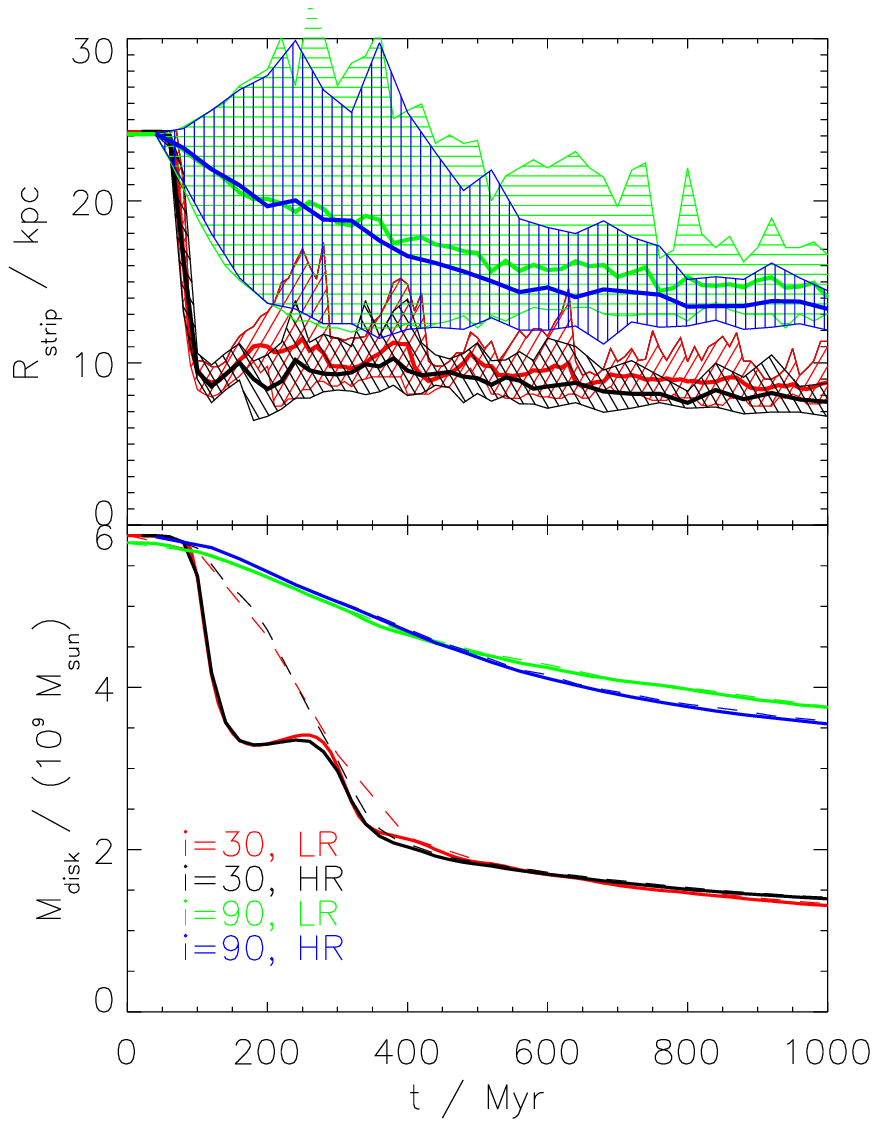}}
\caption{Comparison of disc radius and mass evolution for two different
  resolutions. For medium ram pressure, inclination $30\degree$.}
\label{fig:comp_radius_mass_res}
\end{figure}

In Fig.~\ref{fig:comp_radius_mass_res}, we demonstrate that the disc radius
and mass evolution are nearly independent of resolution. 
%
%
Even the mass loss rate in the continuous stripping phase is very
similar. This is due to the fact that the mass loss is dominated by the
largest KH-modes, which are also resolved in the lower resolution runs.

We conclude that our simulations use a sufficient resolution.

%
\bibliographystyle{mn2e}
\bibliography{%
../../BIBLIOGRAPHY/theory_simulations,%
../../BIBLIOGRAPHY/hydro_processes,%
../../BIBLIOGRAPHY/numerics,%
../../BIBLIOGRAPHY/observations_general,%
../../BIBLIOGRAPHY/observations_clusters,%
../../BIBLIOGRAPHY/observations_galaxies,%
../../BIBLIOGRAPHY/galaxy_model,%
../../BIBLIOGRAPHY/gas_halo,%
../../BIBLIOGRAPHY/icm_conditions,%
../../BIBLIOGRAPHY/else}

\bsp

\label{lastpage}

\end{document}